\documentstyle[12pt]{article}
\normalsize

\def\be{\begin{equation}}
\def\ee{\end{equation}}

\def\bar#1{\overline{#1}}

\def\Hat#1{\rlap{\kern.10em$\widehat{\phantom G}$}#1}
\def\HAt#1{\rlap{\kern.05em$\widehat{\phantom G}$}#1}

\def\cap#1{\rlap{\kern.1em$\widehat{\phantom{G\vrule height.8em}}$}#1{}}
\def\Cap#1{\rlap{\kern.05em$\widehat{\phantom{G\vrule height.8em}}$}#1{}}

\let\oldtheequation=\theequation
\def\doteqs#1{\setcounter{equation}{0}
            \def\theequation{{#1}.\oldtheequation}}
\newcounter{sxn}
\def\sx#1{\addtocounter{sxn}{1} \bigskip\medskip \goodbreak
\noindent{\large\bf
\centerline{\thesxn.~~#1}} \nobreak \medskip}
\def\sxn#1{\sx{#1} \doteqs{\thesxn}}
\newcounter{axn}
\def\ax#1{\addtocounter{axn}{1} \bigskip\medskip \goodbreak
\noindent{\large\bf
{\Alph{axn}.~~#1}} \nobreak \medskip}
\def\axn#1{\ax{#1} \doteqs{\Alph{axn}}}
\newcommand{\ba}{\begin{eqnarray}}
\newcommand{\ea}{\end{eqnarray}}
\def\br{\begin{thebibliography}{abc}}
\def\er{\end{thebibliography}}

\date{}
\begin{document}
\bibliographystyle{unsrt}
\footskip 1.0cm
\thispagestyle{empty}
\setcounter{page}{0}
\begin{flushright}
Alabama UAHEP-9218 \\
G\"oteborg ITP 92-58\\
Napoli DSF-T-92/20,INFN-NA-IV-92/20\\
Rochester UR-1290,ER-40685-739\\
Syracuse SU-4240-528\\
April 1993
\end{flushright}
\begin{center}{\LARGE CURRENT ALGEBRA AND \\
CONFORMAL FIELD THEORY\\
  ON A FIGURE EIGHT\\ }
\vspace*{6mm}
{\large  A. P. Balachandran $^{1}$,
   G. Bimonte $^{1,2}$,
   K. S. Gupta $^{3}$,    \\
   G. Marmo $^{2}$,
          P. Salomonson $^{4}$,
          A. Simoni $^{2}$,
          A. Stern $^{2,5}$ \\ }
\newcommand{\bc}{\begin{center}}
\newcommand{\ec}{\end{center}}
\vspace*{5mm}
 1){\it Department of Physics, Syracuse University,\\
Syracuse, NY 13244-1130, USA}.\\
\vspace*{4mm}
 2){\it Dipartimento di Scienze Fisiche dell' Universit\`a di Napoli,\\
    Mostra d'Oltremare pad. 19, 80125 Napoli, Italy,\\
    and\\
 Istituto Nazionale di Fisica Nucleare, Sezione di Napoli,\\
    Mostra d'Oltremare pad. 19, 80125 Napoli, Italy}.\\
\vspace*{4mm}
 3){\it Department of Physics and Astronomy,\\
  University of Rochester,\\
Rochester, NY 14627, USA}.\\
\vspace*{4mm}
 4){\it Institute of Theoretical Physics,\\
S-41296 G\"oteborg, Sweden}.\\
\vspace*{4mm}
 5){\it Department of Physics, University of Alabama, \\
Tuscaloosa, AL 35487, USA.}\ec

\newpage
\vspace*{5mm}

\normalsize
\centerline{\bf ABSTRACT}

We examine the dynamics of a free massless scalar field
on a figure eight network.
Upon requiring the scalar field to have a well defined value
at the junction of the network, it is seen that the conserved currents
of the theory satisfy Kirchhoff's law, that is that
the current flowing into the junction equals the current flowing out.
We obtain the corresponding current algebra
and show that, unlike on a circle,
the left- and right-moving currents on the figure eight do not
in general commute in quantum theory.
Since a free scalar field theory on a one dimensional spatial
manifold exhibits conformal symmetry, it is natural to ask
whether an analogous symmetry can be defined for
 the figure eight.  We find that, unlike in the case of a manifold,
the action plus boundary conditions for the network are not invariant
under separate conformal transformations associated with left- and
right-movers.  Instead, the system is, at best,
invariant under only a single set of transformations.
Its conserved current is also found to
satisfy Kirchhoff's law at the junction.  We obtain
the associated conserved charges, and
show that they generate a Virasoro algebra.
Its conformal anomaly (central charge)
is computed for special values of the parameters characterizing
the network.
\newpage
\baselineskip=24pt

\sxn{Introduction}

One dimensional networks are simple examples of topological spaces which
are not manifolds.  They can be physically realized
in molecular systems, such as in
the case of polymers, crystals and annulenes\cite{rs,c}.  Also,
the manufacture and study of mesoscopic systems including networks
is of current interest\cite{im}.

In the past, the theory of
networks has been studied in the framework of
quantum mechanics\cite{rs,c,av,bal,an}
 [ as contrasted with quantum field theory ] .
For the case of annulenes, the quantum mechanical particle represents
itinerant $\pi$ electrons which are free to propagate on the network.
Recently, quantum mechanics
was applied to the study of adiabatic transport phenomena\cite{av},
as well as the statistics of identical particles,
on networks\cite{bal,an}.  Topology played a central role in these
studies.


In this article, we shall explore some consequences of defining
a field theory on a network.  Here we choose a simple example of
a field theory, consisting of a single massless scalar field,
and a simple example of a network, the figure eight network.
Physically, we can think of the figure eight as being made
up of two superconducting loops of wire, with the scalar field
representing the order parameter of the superconductor.

The dynamics of massless scalar fields on two dimensional manifolds
(with circle as the spatial slice and the real line
 ${\bf R}^1$ accounting for time) have been well studied.
Free massless scalar fields
 (which we shall also study here) are described by
conformal field theories.  They exhibit the affine $U(1)$ Lie group
 [ the centrally extended loop group $\tilde L U(1)$ of $U(1)$ ]
 and the Virasoro group \cite{go} as
symmetries.  One of the purposes of this investigation is to see
 what happens to these symmetries when the space-time domain is
 figure eight $\times$ ${\bf R}^1$
( ${\bf R}^1$ again accounting for time).

The figure eight consists of two loops with one point
in common, the junction.  For purposes of generality, we shall allow
 the loops to have different lengths,
 $\ell_1$ denoting the length of loop $1$, and
$\ell_2$ denoting the length of loop $2$.  In addition to $\ell_a$,
four other parameters can be used to
characterize a massless free scalar field theory on
a general figure eight network.
They correspond to the velocities of wave propagation
$v_a,$ on loops $a=1$ and $2$, along with the tension,
$T_a,$ or energy per unit length
associated with loop $a$.  In general, the set of values for
$\{\ell_1,\;v_1,\;T_1\} $ may be different from $\{\ell_2,\;v_2,\;T_2\}$.
However, as we shall see in Section 2, the ``physics" of the
figure eight network depends on only four independent combinations
of the parameters $\ell_a,\;v_a$ and $T_a$.

In Section 2, we shall examine the classically conserved currents
associated with a free scalar field theory on the figure eight.  The
  boundary conditions on the fields at the junction
  are crucial in defining the theory.
In this article,  we shall primarily be concerned with
scalar fields which have a well defined value
at the junction, so that they do not possess any discontinuities.
Physically, this is reasonable for a superconducting network (with the
scalar field representing its order parameter), provided a
potential does not exist across the junction.
On the other hand, the associated currents
need not be free of discontinuities.  We find that
the time-component of the current, or charge density,
has a discontinuity at the junction when
${T_1 \over v_1}\ne{T_2 \over v_2}$, while
the space-component of the current must satisfy Kirchhoff's law
which states that
the current flowing into the junction equals the current flowing out.

Section 3 examines the current algebra of the field theory of
Section 2.
As is well known, a quantum field is an operator valued distribution.
The choice of the test function space for such distributions is
an essential part of their definition. Distributions defined on
different test function spaces, in general, are not equivalent.
In this paper, the criterion we follow in order to define the
test function spaces of our fields is that they lead to
well defined Poisson brackets at the classical level. We think that
this is a necessary condition in order to have a consistent
quantization.
We can show that, as a result of this choice of test function
space, the left- and right-moving currents on the figure eight do not
in general commute in quantum theory.  In contrast, the corresponding
currents
 of a free massless scalar field on a manifold do of course commute.

In Section 4, we further study
 the classical currents for two special cases of the
parameters $\ell_a,\;v_a$ and $T_a$
classifying the figure eight.  For the first case (which we refer to
 as case {\bf a}), ${T_1 \over v_1}={T_2 \over v_2}$
and there are no conditions on $\ell_a$, while
for the second case (which we refer to as
case {\bf b}) $\;{\ell_1\over \ell_2}=
{v_1\over v_2}= {T_1\over T_2}$.  The analysis of the currents
simplifies for these cases, as
we obtain certain periodic boundary conditions for the currents in
case {\bf a}, and, even better, periodic currents in case {\bf b}.
The current algebra for the latter case is easily expressible in terms
of three sets of normal modes, and
yields three $U(1)$ current algebras \cite{go} upon quantization.
Two sets of these modes are analogous
to the left- and right-moving modes on a circle, while the remaining
modes are unique to the figure eight.
  We then apply the Sugawara construction to these modes to obtain
three classical Virasoro or Witt algebras with generators we denote by
$L^+_n$, $L^-_n$ and $L^0_{n}$.

Normally, the existence of a Virasoro
algebra indicates the presence of a conformal symmetry.  We
examine the question of conformal symmetry for the figure eight
in Section 5.  We show that, unlike a
massless scalar field theory on a circle, the analogous theory on a
figure eight is not invariant
under separate left and right conformal transformations.  Instead,
the action plus boundary conditions are, at best,
invariant only under a single set of transformations.  The
conserved current corresponding to the conformal symmetry transformation
is shown to satisfy Kirchhoff's
law at the junction.  This conformal symmetry exists provided
${\ell_1 v_2}\over{\ell_2 v_1}$ is rational.  When
${\ell_1 v_2}\over{\ell_2 v_1}$ is not rational, there exists no analogue
of conformal symmetry for the figure eight.  For the former case,
we find the associated conserved charges, and
show that they generate the Virasoro algebra with zero central charge,
which (as alluded to before) is also called the Witt algebra.  If in addition
to ${\ell_1 v_2}\over{\ell_2 v_1}$ being rational,
the parameters satisfy the case {\bf b} conditions
$\;{\ell_1\over \ell_2}= {v_1\over v_2}= {T_1\over T_2}$,
 this algebra is spanned by
$L^+_n+L^-_n+{1\over 2}L^0_{2n}$.
Until this stage, our treatment is purely classical.
The quantum mechanical version of the above algebra,
complete with the central extension, is commented on at the end
of Section 5.

In Section 6, we show that, unlike on a circle, the left- and
right-moving chiral currents of the classical theory
cannot be independently quantized on the figure eight.
By this we mean that the two chiral currents cannot be expanded
 in terms of two independent sets of bases such that their quantum
 analogues i) have a well defined action on the Fock space, and
ii) provide a quantization of the currents
which is unitarily equivalent to that derived
from the eigenmodes of the one-particle Hamiltonian of the system.

In Appendix A of this paper,
we sketch the possibility of having discontinuous boundary conditions
for the scalar field at the junction.  Boundary conditions, in general,
are restricted only by the requirement that a certain differential
 operator acting on a Hilbert space of
square integrable functions is self-adjoint, and there are such
conditions admitting these discontinuities.
  In Appendix B,
we write down the general solutions to the field equations on the
figure eight consistent with the boundary conditions of Section 2,
and carry out the eigenmode expansions of fields and currents
for two special choices of the parameters of the figure eight
corresponding to cases {\bf a} and {\bf b}.

\sxn{ The Singlevaluedness Condition and Kirchhoff's Law}

We first introduce a set of coordinates on the figure
eight.  Let $x$ be the spatial coordinate, with $0 \le x \le
\ell_1+\ell_2$,
and let $t$ be time.  We choose $x$ so that we are on loop $1$
when $0 \le x \le \ell_1$ and we are
on loop $2$ when $\ell_1 \le x \le \ell_1+\ell_2\;$.
 $x=0=\ell_1=\ell_1+\ell_2$ are all assumed
to correspond to the same point, namely the junction (see Figure 1).
Next, we introduce a complex scalar field $\Phi$ which is a function
of $x$ and $t$.
For the sake of simplicity, let us hold the magnitude of $\Phi(x,t)$
to be fixed at one, so that it just defines a single degree of freedom,
a phase.  If desired, we can justify this approximation by
assuming the presence of a symmetry breaking
potential in the Lagrangian for the system,
such as $V(\Phi)=\lambda(\Phi^*\Phi-1)^2$.  Then $\Phi^*\Phi$ is frozen to $1$
and we are
left with just a phase $\chi$ defined by
\be
\Phi (x,t)=e^{i\chi (x,t)}\;
\ee
in the limit $\lambda \rightarrow \infty$. For the dynamics of $\chi(x,t)$,
we shall assume the free wave equation
\be
\Bigl[
\; \partial_x^2 - {1\over{v^2_a}}\partial_t^2\; \Bigr]
\;\chi(x,t) = 0
\ee
where $v_a$ represents the wave velocity on loop $a$.

\input{psbox.tex}

\begin{figure}[hbtp]
\begin{center}
\mbox{\psannotate{\psboxto(11cm;0cm){figure.eps}}
{\at(9\pscm;16\pscm){$x=0=l_1=l_1+l_2$}
\at(3.5\pscm;7\pscm){$0 \le x \le l_1$}
\at(15.5\pscm;7\pscm){$ l_1 \le x \le l_1+l_2$}
\at(5\pscm;1\pscm){$loop~1$}
\at(17\pscm;1\pscm){$loop~2$} }}
\end{center}
\begin{center}
{\footnotesize{\bf Fig. 1.} Figure 8 with its coordinates.}
\end{center}
\end{figure}

Rather than work with the spatial coordinate
$x$, we find it more convenient to use another
coordinate $\sigma $ where $0\le\sigma\le 2\pi$.
It is defined so that
there is a two-to-one mapping from $\{x\}$ to $\{\sigma\}$.
It is such that, a given value of $\sigma$ corresponds to a point
on loop 1, and also to a point on loop 2.
The relation between $x$ and $\sigma$ for points on loop $1$ is
\be
x={{\ell_1}\over{2\pi}}\sigma \;,
\ee
while for loop $2$, it is
\be
x={{\ell_2}\over{2\pi}}\sigma +\ell_1\;.
\ee

With $\sigma$ as the coordinate, it becomes necessary
to distinguish the fields on the two loops of figure eight.
For this purpose, we replace $\Phi$ by a two component field
$\phi$ where $\phi(\sigma,t)=
(\phi_1(\sigma,t),\phi_2(\sigma,t))$.
$\phi_a$ corresponds to the field $\Phi$ evaluated
on loop $a$.  More precisely, we define $\phi_a$ by
\be
\phi_1(\sigma,t)= \Phi\Bigl({{\ell_1}\over{2\pi}}\sigma,t\Bigr)
\quad {\rm and} \quad
\phi_2(\sigma,t)= \Phi\Bigl({{\ell_2}\over{2\pi}}\sigma +\ell_1,
t\Bigr)  \quad .
\ee

Since $\Phi$ is a phase, so is $\phi_a$ and we can write
$\phi_a (\sigma,t)=e^{i\chi_a (\sigma,t)}\;.$
In terms of the degrees of freedom $\chi_a$, the wave equation (2.2)
becomes
\be
\Bigl[ {1\over{\kappa_a^2}}\partial_\sigma^2 - \partial_t^2 \Bigr]
\;\chi_a(\sigma,t) = 0 \;,\quad   a=1,2\;,
\ee
where
$\kappa_a={{\ell_a}\over{2\pi v_a}} $.

Eq. (2.6), by itself, is not sufficient to completely specify the
dynamics of the system.  It has to be supplemented with boundary
conditions on the fields $\chi_a$ at the junction.
To show this, we first note that the substitution of
$\chi_a (\sigma,t)=e^{i\omega t}\tilde\chi_a(\sigma)$
in eq. (2.6) leads to the eigenvalue equation
\be
\Bigl [ H_a -\omega^2 \Bigr ]   \;\tilde\chi_a(\sigma) = 0 \;,\qquad
H_a=- {1\over{\kappa_a^2}}\partial_\sigma^2 \;.
\ee
The eigenfunctions of $H_a$ will be interpreted as single particle
wavefunctions in quantum theory.
These eigenfunctions
must form a complete set in the Hilbert space of square integrable
functions of the figure eight for time evolution
of the second quantized field theory to be unitary.
This space consists of
functions $\tilde\chi \equiv (\tilde\chi_1,\tilde\chi_2)$ with the
inner product
\be
<\tilde\chi,\;\tilde\psi>=\sum_{a=1,2}
\nu_a \kappa_a^2   \int  d\sigma\;
\tilde\chi_a^{*}(\sigma)\tilde\psi_a(\sigma)\;.
\ee
Here we have introduced a new parameter $\nu_a$, since in general,
the coefficient of the measure $d\sigma$
 need not be the same for the two loops.

Let us define the single particle Hamiltonian $H$ by
$H\tilde\chi=(H_1\tilde\chi_1,H_2\tilde\chi_2)$.
This is only a formal definition of $H$, since
we have not specified its domain ${\cal D}(H)$.
We can require that ${\cal D}(H)$ is so chosen
that $H$ is self-adjoint, this condition being compatible
with physical principles.  There are an infinite number of
options for ${\cal D}(H)$ consistent with this requirement.  They
 correspond to different
boundary conditions for the functions $\tilde\chi\in
{\cal D}(H)$ at the junction, and lead to inequivalent definitions
of the operator $H$.
(We sketch the different possibilities in Appendix A.)
These possibilities can be thought of as describing different
junctions and the right one, in a given problem,
must be chosen on physical grounds.

In this paper, we think of the figure eight as made of two
superconducting wires, and of the field $\Phi$
as the order parameter.  Then,
if no potential is applied across the junction, $\Phi$ has to be
continuous there.  Therefore,
\be
\Phi(0,t)=\Phi (\ell_1,t)=\Phi(\ell_1+\ell_2,t)\;,
\ee
or in terms of $\phi_a$,
\be
\phi_1(0,t)=\phi_1 (2 \pi,t)=\phi_2(0,t)=\phi_2(2 \pi,t)\;.
\ee
Consequently, $\chi$ is allowed to have $2 \pi$ discontinuities
across the junction. These discontinuities, which represent
winding modes of the field $\phi$ or $\chi$, are topologically stable
under time evolution.  A typical winding mode for $\chi_a$ is
proportional to $\sigma$ for all $t$.  It fulfills the wave
equation (2.6).
After subtracting such modes from $\chi_a$, we can regard
$\chi$ too to be continuous across the junction.
This requirement picks up a unique definition for $H$, namely that
specified by the following domain:
\be
{\cal D}(H)=
 \biggl\{~\tilde\chi~|~ \tilde\chi_1(0)=\tilde\chi_1 (2\pi)
 =\tilde\chi_2(0)=\tilde\chi_2(2 \pi)\;,
{}~\sum_{a=1,2} \nu_a \partial_{\sigma} \tilde
\chi_a |_0^{2 \pi}=0~ \biggr\}.
\ee

It is easy to verify that $H$ defined above is self-adjoint in the
following manner:
If $\tilde\chi$ is an arbitrary element of ${\cal D}(H)$, then the
domain  ${\cal D}(H^{\dagger})$ of the adjoint $H^{\dagger}$ of $H$
consists of functions $\tilde \psi$ in the Hilbert space which fulfill
\be
<\tilde\psi,\;H \tilde\chi> =<H \tilde\psi ,\;\tilde\chi> \;.
\ee
$H$ is self-adjoint if and only if ${\cal D}(H^{\dagger})={\cal D}(H)$.
Now eq. (2.12) implies that
\begin{eqnarray}
0&=&-\sum_{a=1,2} \nu_a\int d\sigma\;
\tilde\psi^*_{ a}(\sigma) \partial_\sigma^2 \tilde\chi_{a}(\sigma)
 +\sum_{a=1,2} \nu_a \int d\sigma\;
\partial_\sigma^2 \tilde\psi^*_{a}(\sigma) \tilde\chi_{a}(\sigma)
\nonumber\\
&=&\sum_{a=1,2} \nu_a \biggl(\partial_{\sigma}
\tilde \psi^*_{a} \tilde \chi_{a} -
 \tilde \psi^*_{a}
 \partial_{\sigma} \tilde \chi_{a}\biggl)\bigg|_0^{2 \pi}\\
&=&\biggl(\sum_{a=1,2} \nu_a \partial_{\sigma}
\tilde \psi^*_{a} |_0^{2 \pi} \biggr)
\tilde \chi_{1} |_0 + \sum_{a=1,2}
\biggl( \tilde\psi^*_{a}|_0 -\tilde\psi^*_2|_{2\pi}\biggr)\nu_a
 \partial_{\sigma} \tilde\chi_{a}|_0
-\biggl( \tilde\psi^*_{1}|_{2\pi} -\tilde\psi^*_2|_{2\pi}\biggr)\nu_1
 \partial_{\sigma} \tilde\chi_{1}|_{2\pi}   \;\; \;.   \nonumber
\end{eqnarray}
Since the boundary values of $\tilde\chi$ and
$\partial_{\sigma} \tilde\chi$ are arbitrary
but for the conditions (2.11), we must have
$$
\tilde\psi_1(0)=\tilde\psi_1 (2 \pi)
 =\tilde\psi_2(0)=\tilde\psi_2(2 \pi)\quad {\rm and  }\quad
\sum_{a=1,2} \nu_a \partial_{\sigma} \tilde\psi_a |_0^{2 \pi}=0  \;.
$$
Hence $\tilde\psi$ is an element of ${\cal D}(H)$, or equivalently,
the domain ${\cal D}(H^{\dagger})$
of $H^{\dagger}$ is the same as ${\cal D}(H)$,
 proving that $H$ is self-adjoint.

The wave equation (2.2) and the boundary conditions (2.11) are
obtainable from an action principle,
the action $S$ being the sum of two terms:
\be
S=S_1+S_2\;,\quad
S_a={{\nu_a}\over 2}\int d\sigma dt\;\Bigl(
{\kappa_a^2}|\partial_t \phi_a|^2
- |\partial_\sigma \phi_a|^2\Bigr)\;.
\ee
In the original coordinates $(x,t)$, the terms
$S_1$ and $S_2$ can be written as
$$
S_1={T_1\over 2}\int dt \int_0^{\ell_1} dx\;\Bigl(
{1\over{v_1^2}}|\partial_t \Phi|^2
- |\partial_x \Phi|^2\Bigr)\;,
$$
\be
S_2={T_2\over 2}\int dt \int_{\ell_1}^{\ell_2} dx\;\Bigl(
{1\over{v_2^2}}|\partial_t \Phi|^2
- |\partial_x \Phi|^2\Bigr)\;,
\ee
$$
{}~T_a:= \frac {\nu_a l_a}{2 \pi}\;.
$$
{}From (2.15), we see that $T_a$ can be interpreted as the
``tension'' in loop $a$.  If the loops are made of
different superconducting
materials, there is no reason why $T_a$ should be identical.

To obtain the wave equation (2.6), we extremize (2.14)
for variations of the field $\phi_a$ which vanish at the junction.
If next we allow also variations of $\chi$
that are continuous at the junction,
we recover in addition the boundary condition
$\sum_{a=1,2} \nu_a \partial_{\sigma} \chi_a |_0^{2 \pi}=0.$

The solutions to the equations of motion (2.6) are of the form
\be
\chi_a(\sigma,t) =
\chi_a^+(\sigma_a^+) + \chi_a^-(\sigma_a^-) \quad,
\ee
where $\sigma_a^\pm =\kappa_a\sigma\pm t$.

The equations of motion (2.6) can be recast in terms of current
conservation laws.
For this purpose, we define the time-components of the currents by
\be
 J_t^a=-{{i\nu_a\kappa_a}\over 2}(\phi^*_a\partial_t\phi_a -
\phi_a\partial_t\phi^*_a)=\nu_a\kappa_a\partial_t\chi_a
 \qquad{\rm (no\; sum\; on\;} a)\quad,
\ee
and the space-components by
\be
J_\sigma^a=-{{i \nu_a}\over {2}}(\phi^*_a\partial_\sigma\phi_a -
\phi_a\partial_\sigma\phi^*_a) =  \nu_a\partial_\sigma \chi_a
\qquad{\rm (no\; sum\; on\;} a)\quad.
\ee
Then eqs. (2.6) imply that the currents are conserved:
\be
 \kappa_a\partial_t J_t^a - \partial_\sigma J_\sigma^a= 0\;.
\ee

{}From the solutions (2.16) to the equations of motion,
we can form left- and
right- moving combinations $J_\pm^a$ of currents.  They are defined
according to
\be
J_\pm^a =  J_\sigma^a \pm J_t^a\;.
\ee
The solutions imply that $J_+^a$ and $J_-^a$ is each a function of
just one variable:
\be
 J_+^a(\sigma_a^+)=2\nu_a\kappa_a\;{{\partial\chi_a^+}\over
{\partial\sigma_a^+}} \quad {\rm and}\quad
 J_-^a(\sigma^-_a)=2\nu_a\kappa_a\;{{\partial \chi_a^-}\over
 {\partial\sigma^-_a}} \quad.
\ee

Our choice for the coordinate $\sigma$ selects a particular
orientation on the figure eight.
The space-component of currents $ J_\sigma^a$ will be regarded as positive
(negative) when the current flows in the direction of increasing
(decreasing) $\sigma$.
Thus a positive $ J_\sigma^a(0,t)$ corresponds to a current
leaving the junction and flowing into loop $a$,
and a positive $ J_\sigma^a(2\pi,t)$
corresponds to a current entering the junction from loop $a$.
The boundary condition for the space derivatives of $\chi$
is therefore just the Kirchhoff law for the currents,
as it states that the total current flowing into the junction equals
the total current flowing out of the junction:
\be
\sum_{a=1,2} J_\sigma^a(0,t)
=\sum_{a=1,2} J_\sigma^a(2\pi,t) \quad.
\ee
By taking the time derivative of this condition, we further have that
\be
{d\over {dt}}\biggl(
\sum_{a=1,2} \nu_a \partial_\sigma \chi_a \biggr)\bigg|^{2\pi}_0 =0
{}~~{\rm or}~~ \biggl( \sum_{a=1,2} \frac{1}{\kappa_a} \partial_{\sigma} J_t^a
\biggr) \bigg| ^{2 \pi}_0 =0.
\ee

Boundary conditions exist also for the time-components
$J^a_t$ of the currents.  They are obtained by
requiring that the boundary conditions (2.10) are preserved in time, that
is $\partial_t\phi_1(0,t)=\partial_t\phi_1(2\pi,t)=$
$\partial_t\phi_2(0,t)=\partial_t\phi_2(2\pi,t)$.
This implies that the time component of the current satisfies
\be
{1\over{\nu_1\kappa_1}}J_t^1(0,t)=
{1\over{\nu_1\kappa_1}}J_t^1(2\pi,t)=
{1\over{\nu_2\kappa_2}}J_t^2(0,t)=
{1\over{\nu_2\kappa_2}}J_t^2(2\pi,t)\;.
\ee
Thus the charge density is discontinuous at the junction when
$\nu_1\kappa_1 \ne\nu_2\kappa_2$.

\sxn{ Current Algebra}

In the Hamiltonian formulation of the theory, $\kappa_a J^a_t$
is canonically conjugate to the field $\chi_a$.
We thus have the equal time Poisson brackets
\be
\{\chi_a(\sigma,t),J^b_t(\sigma',t)\}=
{1 \over { \kappa_a}}\delta^b_a\;\delta(\sigma-\sigma')\;.
\ee
Upon using the definition (2.20) for the left and right moving currents
$J_\pm^a$, we can also $naively$ compute the Poisson brackets between the
currents:
\begin{eqnarray}
\{J^a_\pm(\sigma,t),J^b_\pm(\sigma',t)\}
&=&\pm{{2\nu_a}\over{\kappa_a}}
\delta^b_a\;\partial_\sigma\delta(\sigma-\sigma')\;,   \\
\{J^a_+(\sigma,t),J^b_-(\sigma',t)\}&=&0\;.
\end{eqnarray}

This resembles the current algebra for two scalar fields on a
circle.  However,
the results (3.1-3) are only formal because we have not a) defined
the delta function on a figure eight, and b) taken into
account the boundary conditions on the fields and currents.  Thus, for
instance, the Poisson brackets which we have found cannot be valid in
the limit where we approach the junction $\sigma=0,2\pi$.  As a result,
 the application of (3.1-3) leads to incorrect
  Hamilton's equations of motion
 for the time evolution of the system, as
shown by the following:  From the Hamiltonian
\be
{\cal H}=\sum_{a=1,2}{1\over{4\nu_a}}\int_0^{2\pi}d\sigma
\Bigl[ \Bigl(J_+^a(\sigma)\Bigr)^2 + \Bigl(J_-^a(\sigma)\Bigr)^2  \Bigr]
\quad,
\ee
and (3.3), one would obtain the result that
$\int_0^{2\pi}d\sigma  \Bigl(J_+^a(\sigma)\Bigr)^2 $ is a constant of
 motion:
\be
\partial_t\int_0^{2\pi}d\sigma  \Bigl(J_+^a(\sigma)\Bigr)^2  =
\Bigl\{\int_0^{2\pi}d\sigma  \Bigl(J_+^a(\sigma)\Bigr)^2\;,\;{\cal H}
\;\Bigr\} \;\;=0    \quad.
\ee
But this is incorrect because
from current conservation (2.19) and the identity
$\partial_\sigma J^a_t =\kappa_a\partial_t J^a_\sigma$,
we instead get
\be
\partial_t\int_0^{2\pi}d\sigma  \Bigl(J_+^a(\sigma)\Bigr)^2  =
{1\over {\kappa_a}}  \Bigl(J_+^a(\sigma)\Bigr)^2\Big|_{0}^{2\pi}
 \quad,
\ee
where $  \Bigl(J_+^a(\sigma)\Bigr)^2  $ need not satisfy $2\pi$
periodic boundary conditions.

In order to account for the boundary
conditions and obtain the correct Poisson bracket relations,
let us introduce a set of ``smearing" or ``test" functions
$\Lambda=(\Lambda_+^1,\Lambda_+^2,\Lambda_-^1,\Lambda_-^2)$
 for the currents $J_\pm^a$.
(We shall ignore the $t$ dependence.)
Next we define the ``smeared current" ${\cal J}(\Lambda)$ as follows:
\begin{eqnarray}
{\cal J}(\Lambda)&=&\sum_{a=1,2}\int d\sigma\;
[\Lambda_+^a (\sigma)J_+^a(\sigma)  + \Lambda_-^a (\sigma)J_-^a(\sigma)]
\nonumber\\
&=&\sum_{a=1,2}\int d\sigma\; [(\Lambda_+^a+\Lambda_-^a)
\nu_a\partial_\sigma \chi_a +(\Lambda_+^a-\Lambda_-^a)J^a_t ]\quad.
\end{eqnarray}
In order to be able to define Poisson brackets involving the ``smeared current"
${\cal J}(\Lambda)$ consistently,
 we shall require that ${\cal J}(\Lambda)$
is differentiable with respect to
the phase space variables $\chi_a$\cite{Tr} and $ J^a_t$.
{}From the definition (3.7), we see that
differentiability is assured for variations in
$ J^a_t$.  But that is not, in general, true for variations
$\delta\chi_a$ in
$\chi_a$ as such variations will in general create
boundary terms:
\be
\delta{\cal J}(\Lambda)=
-\sum_{a=1,2}\nu_a\int d\sigma\;\partial_\sigma (\Lambda_+^a+\Lambda_-^a)
\;\delta\chi_a +
\sum_{a=1,2}\nu_a\; (\Lambda_+^a+\Lambda_-^a)\;\delta\chi_a
\bigg|^{\sigma=2\pi}_{\sigma=0}   \quad.
\ee
If we assume continuity of the phase at the junction, so that
$ \delta\chi_1(0,t)=$   $\delta\chi_1(2\pi,t)
=$      $\delta\chi_2(0,t)=$        $\delta\chi_2(2\pi,t)$,
then the boundary term in eq. (3.8)
can be made to vanish by requiring that
\be
\sum_{a=1,2}\nu_a\; (\Lambda_+^a+\Lambda_-^a)
\bigg|^{\sigma=2\pi}_{\sigma=0}=0   \quad.
\ee

We call ${\cal T}$ the space of all test functions
$\Lambda$ satisfying  eq. (3.9).  It is our test function space
for the currents.  For $\Lambda\in{\cal T}$,
the variational derivatives of ${\cal J}(\Lambda)$ with respect to
$\chi_a(\sigma)$ and $ J^a_t(\sigma)$ are given by
\be
{{\delta{\cal J}(\Lambda)}\over{\delta \chi_a(\sigma)}}=
-\nu_a\partial_\sigma
\biggl[\Lambda_+^a(\sigma)+\Lambda_-^a(\sigma)\biggr]  \quad {\rm and}\qquad
{{\delta{\cal J}(\Lambda)}\over{\delta J^a_t(\sigma)}}=
\Lambda_+^a(\sigma)-\Lambda_-^a(\sigma)\qquad .
\ee
\indent
We can now compute the Poisson brackets between two smeared currents
${\cal J}(\Lambda)$ and ${\cal J}(\bar\Lambda)$,
for $\Lambda,\bar \Lambda\in{\cal T}$. Care is necessary in performing
this computation as the $\delta$ functions in (3.1) and (3.2) do not have
all the usual properties. Perhaps the best way is to start with the basic
definition
\be
\{{\cal J}(\Lambda),{\cal J}(\bar\Lambda)\}~=~\sum_{a=1,2}
\frac{1}{\kappa_a}
\int_0^{2 \pi} d \sigma~\biggl\{
{{\delta{\cal J}(\Lambda)}\over{\delta \chi_a(\sigma)}}
{}~{{\delta{\cal J}(\bar\Lambda)}\over{\delta J^a_t(\sigma)}}~-~(\Lambda
\leftrightarrow \bar\Lambda)\biggr\}
\ee
of the Poisson bracket and use (3.10). We then find,
$$
\{{\cal J}(\Lambda),{\cal J}(\bar\Lambda)\}=
\sum_{a=1,2}{{\nu_a}\over{\kappa_a}}
\int d\sigma (\Lambda^a_+ \partial_\sigma \bar\Lambda^a_+
-\bar\Lambda^a_+ \partial_\sigma\Lambda^a_+)
-\sum_{a=1,2}{{\nu_a}\over{\kappa_a}}
\int d\sigma (\Lambda^a_- \partial_\sigma \bar\Lambda^a_-
-\bar\Lambda^a_- \partial_\sigma\Lambda^a_-)
$$
\be
+\sum_{a=1,2}{{\nu_a}\over{\kappa_a}}
(\Lambda^a_+  \bar\Lambda^a_- -\bar\Lambda^a_+ \Lambda^a_-)
\bigg|^{\sigma=2\pi}_{\sigma=0}   \quad.
\ee
Of special interest is the boundary term in eq. (3.12).
It is zero when the
smearing functions are continuous at the junction, so that
$\Lambda_\pm^a(2\pi)=\Lambda_\pm^a(0)$ and
$\bar\Lambda_\pm^a(2\pi)=\bar\Lambda_\pm^a(0)$.
In that case, we recover the
result that the Poisson brackets between left and right moving currents
are zero, which is consistent with eq. (3.3).  On the other hand,
eq. (3.12) shows that, contrary to eq. (3.3), there may be cases where
the Poisson brackets between left and right moving currents
do not vanish.  This can happen, for instance,
when one of the test functions $\Lambda_+^a$ or
$\bar\Lambda_-^a$ is not continuous at the junction
[but consistent with eq. (3.9)].

\sxn{ Periodic Boundary Conditions and Currents}

We now examine the boundary conditions on the currents for
two special cases of the parameters $\kappa_a$ and $\nu_a$.
They are: ${\bf a.}\; \kappa_1 \nu_1  =\kappa_2\nu_2$ (or equivalently,
${T_1 \over v_1}={T_2 \over v_2}$), and
${\bf b.} \;\kappa_1=\kappa_2=\kappa\;$ and $\;\nu_1=\nu_2=\nu\;$.
As the case ${\bf b}$ parameters satisfy
$\kappa_1 \nu_1  =\kappa_2\nu_2$, case
 ${\bf b}$ is actually a subcase of case ${\bf a}$.
(A detailed analysis of the solutions to the equations of motion (2.6)
for both of these cases will be discussed in Appendix B.)

We first consider case {\bf a}.

Case ${\bf a}  \quad \kappa_1 \nu_1  =\kappa_2\nu_2$

Here we can show that the currents $J^a_\pm(\sigma, t)$ can be
 written as linear combinations of functions of $\sigma$ and $t$,
where these functions satisfy
 either $2\pi$ or $4\pi$ periodic boundary conditions with regards to
the spatial coordinate $\sigma$.

Functions with $2\pi$ periodic boundary conditions are obtained
by taking the sum of $J^1_\pm$ and $J^2_\pm$:
\be
J^{sum}_\pm :=  J^1_\pm+J^2_\pm  \;,\quad
 J^{sum}_\pm (2\pi,t) =J^{sum}_\pm (0,t) \;.
\ee
This result is due to the Kirchhoff law (2.22)
and the boundary conditions (2.24), which reduce to
$ J_t^1(0,t)=  J_t^1(2\pi,t)=
J_t^2(0,t)= J_t^2(2\pi,t)\; $ when  we set
$ \kappa_1 \nu_1  =\kappa_2\nu_2$.

Functions with $4\pi$ periodic boundary conditions can be constructed
by first taking the difference of $J^1_\pm$ and $J^2_\pm$:
$$ J_\pm^{dif} : =   J^2_\pm-J^1_\pm \;.$$
 Then eqs. (2.22) and (2.24) imply that
\be
 J^{dif}_+(0,t)= J^{dif}_-(0,t) \quad  {\rm and} \quad
 J^{dif}_+(2\pi,t)= J^{dif}_-(2\pi,t)  \;.
 \ee
To analyze these conditions it is helpful to
introduce yet another function ${ K(s,t)}$, which
 is defined on the spatial domain  $\{s;\;0\le s\le 4\pi\}$ as follows:
\be
  K(s,t)=\left\{\matrix{ J^{dif}_+(s,t) \;, \quad {\rm if}
\quad 0 \le s\le 2\pi  \cr
 J^{dif}_-(4\pi-s,t) \;,
\quad {\rm if}\quad 2\pi \le s\le 4\pi  \cr } \right.   \;.
\ee
In view of  eqs. (4.2),
this function is continuous in $s$ and satisfies the
$4\pi$ periodic boundary condition $  K(0,t) =  K(4\pi,t) $.

Case ${\bf b}. \quad
\kappa_1=\kappa_2=\kappa\;$ and $\;\nu_1=\nu_2=\nu\;$

We discuss this case in the remainder of this Section.

In case ${\bf b}$,
the time evolution of the functions $J^{sum}_\pm(\sigma,t)$
  and $ K(s,t)$ can be given in a simple closed form, analogous
to that found for the chiral currents on a circle.  In fact, they
can be expressed in terms of periodic functions of only
one argument.  We denote these functions by $f^{sum}_\pm$ and $f$.
Then the result may be stated as follows:
\begin{eqnarray}
 J^{sum}_\pm(\sigma,t)~&=&~f^{sum}_\pm(\sigma\pm t/\kappa)   \;,
 \nonumber\\  K(s,t)~&=&~f(s+ t/\kappa)   \;,
\end{eqnarray}
where
\be
f^{sum}_\pm(x + 2 \pi)~=~f^{sum}_\pm(x)~~,~~ \quad
f(x+ 4 \pi)~=~f(x)~~;~~ -\infty < x < \infty    \;.
\ee

To prove eqs. (4.4) and (4.5), we just recall that, as a consequence of
eq. (2.16) and thanks to the condition $\kappa_1=\kappa_2=\kappa$, we can
write the currents $J_\pm^a$ as functions of just a single variable, the
same for both loops as in eq. (2.21):
\be
J_\pm^a(\sigma,t)=f_\pm^a(\sigma\pm t/\kappa)\;.
 \ee
 Upon substituting (4.6) into (4.1), we get
$$f^{sum}_\pm(2\pi\pm t/\kappa)=f^{sum}_\pm(\pm t/\kappa)$$
where $f^{sum}_\pm(x) \equiv (f_\pm^1+f_\pm^2)(x)$. This
is equivalent to the result $f^{sum}_\pm(x + 2 \pi)~=~f^{sum}_\pm(x)~~$.

As for the function $K(s,t)$, upon substituting (4.6) into (4.3), we get
\be
  K(s,t)=\left\{\matrix{
\biggl(f_+^2-f_+^1\biggr)(s+t/\kappa) \equiv f^{dif}_+ (s+t/\kappa)
   \;, \quad {\rm if} \quad 0 \le s\le 2\pi  \cr
\biggl(f_-^2-f_-^1\biggr)(4\pi-s-t/\kappa) \equiv f^{dif}_-(s+t/\kappa)~,
\quad {\rm if}\quad 2\pi \le s\le 4\pi  \cr } \right.   \;.
\ee
But we have already proved, under case {\bf a},
that the function $K(s,t)$
is continuous in $s$ (in particular, at $s=2 \pi$)
and satisfies $4 \pi$ periodic conditions at all times.
The former implies that
$$
 f^{dif}_+ (2\pi+t/\kappa) =  f^{dif}_-(2\pi+t/\kappa)~,
$$
or that $ f^{dif}_+ (x) =  f^{dif}_-(x)\equiv f(x)~,$ while the latter
implies that
$$
 f^{dif}_+ (t/\kappa) =  f^{dif}_-(4\pi+t/\kappa)~,
$$
or that $f(4\pi + x) = f(x)$.  We have thus proved eqs. (4.4)
and (4.5).

The periodicity of the currents
allows us to make the Fourier expansions
\begin{eqnarray}
 J^{sum}_\pm(\sigma,t)&=&
\mp\sum_{n=-\infty}^\infty \alpha^\pm_n(0)\; e^{-in(t/\kappa \pm
\sigma)}\quad, \\
 K(s,t)&=&-\sum_{n=-\infty}^\infty \beta_{n\over 2}(0)\;e^{-\frac{in}{2}
(t/\kappa + s) }\quad,
\end{eqnarray}
where $n=0,\pm1,\pm2,...\;.$ and
$\alpha^\pm_n(0) \equiv \alpha^\pm_n$ and $\beta_{n\over 2}(0)
\equiv\beta_{n\over 2}$
represent the values of the coefficients at time $t=0$.
The reality of the currents implies that
$\alpha^\pm_{-n}=(\alpha^\pm_{n})^*$
and $\beta_{-{ n\over 2}}=(\beta_{n\over 2})^*$.
{}From eqs. (4.8) and (4.9), we can obtain a basis for the test functions
$\Lambda=(\Lambda_+^1, \Lambda_+^2,\Lambda_-^1,\Lambda_-^2)$
appearing in the smeared currents (3.7).
The test functions associated with the coefficients
$\alpha^+_n$, $\alpha^-_n$ and $\beta_{n\over 2}$ are
\begin{eqnarray}
\Lambda^{(\alpha^+_n)}&=&- {1\over {2\pi} }  \;
(e^{in\sigma},\; e^{in\sigma},\;0,\; 0)\;,\nonumber \\
\Lambda^{(\alpha^-_n)}&=&  {1\over {2\pi}}   \;
(0,\;0,\;e^{-in\sigma},\; e^{-in\sigma})\; ,   \\
\Lambda^{(\beta_{n\over 2})}&= &{1\over {4\pi} }  \;
(e^{in\sigma / 2},\;-e^{in\sigma / 2},\;
e^{-in\sigma / 2},\;-e^{-in\sigma / 2}) \;\;, \nonumber
\end{eqnarray}
respectively.  These test functions satisfy the condition (3.9),
and hence belong to the set ${\cal T}$.
As $\alpha^+_n$, $\alpha^-_n$ and $\beta_{n\over 2}$ form a
complete set of coefficients, the $\Lambda^{(X)}$'s for
$X=\alpha^+_n$, $\alpha^-_n$ and $\beta_{n\over 2}$ form a
complete set of test functions spanning ${\cal T}$.
The first two types of test functions
$\Lambda^{(  \alpha^\pm_n)}$ are associated
with left- and right-moving modes, analogous
to the modes on a circle, while the last type of test functions
$\Lambda^{( \beta_{n\over 2})}$ is unique to the figure eight.

The above $\Lambda^{(X)}$'s, are normalized to satisfy
\be
{\cal J}(\Lambda^{(X)})=X\;.
\ee
We can use this relation and (3.12)
to compute the Poisson brackets of
$\alpha^\pm_n$ and $\beta_{n\over 2}$.  The nonzero brackets
\be
\{\alpha^\pm_{m},\;\alpha^\pm_{-n}\}=
- {{2 in \nu }\over {\pi\kappa}}   \;\delta_{m,n}     \;,\quad
\{\beta_{m \over 2},\;\beta_{-{n \over 2}}\}=
 -{{in\nu }\over {2\pi\kappa}}   \;\delta_{m,n}  \;,
 \ee
define three U(1) affine Lie algebras \cite{go}.

Of course, given the three classical affine
U(1) algebras above, we can construct
three classical Virasoro or Witt algebras, the generators being
\be
L^\pm_n= {{\pi\kappa}\over{4\nu}}\sum_m \alpha^\pm_m\alpha^\pm_{n-m}
\quad{\rm and}\quad
L^0_n={{\pi\kappa}\over{\nu}}
\sum_m \beta_{m \over 2}\beta_{{n-m}\over 2} \;.
\ee
{}From the Poisson brackets (4.12), it follows that
\be
\{L^\pm_n\;,\;L^\pm_m\}=-i(n-m)L^\pm_{n+m}
\quad{\rm and}\quad
\{L^0_n\;,\;L^0_m\}=-i(n-m)L^0_{n+m} \;.
\ee

Just as for conformal field theories on a circle,
the $n=0$ generators appear in
the expression for the Hamiltonian since
\begin{eqnarray}
{\cal H}&=&{1\over{4\nu}}  \sum_{a=1,2}  \int_0^{2\pi}d\sigma
\Bigl[ \Bigl(J_+^a(\sigma,t)\Bigr)^2 + \Bigl(J_-^a(\sigma,t)\Bigr)^2  \Bigr]
\nonumber\\
&=&{1\over{\kappa}}(L^+_0 + L^-_0 +{1\over 2}L^0_0)  \;.
 \end{eqnarray}

In quantum theory, we promote $\alpha^\pm_n$ and $\beta_{n\over 2}$
to operators, and replace the Poisson brackets of (4.12)
by $-i$ times commutator brackets.  The quantum
operators act on a Fock space, and we assume, as usual,
that $\alpha^\pm_n$ and $\beta_{n\over 2}$
for $n>0$ annihilate its vacuum $|0>$ and are destruction operators.
The nonvacuum states of the Fock space
are obtained by acting on $|0>$ with
$\alpha^\pm_n$ and $\beta_{n\over 2}$ for $n \le 0$.
The quantum version of the Virasoro generators are assumed to
be normal ordered, with destruction operators appearing on the
right.  The classical Virasoro algebras are then modified by the
standard central terms, with each algebra having central charge
$c=1$:
$$
[ L^\pm_n\;,\;L^\pm_m ] =(n-m)L^\pm_{n+m}
+{1\over {12}}n(n^2-1)\delta_{n+m,0}
\quad ,
$$
\be
[ L^0_n\;,\;L^0_m ] =(n-m)L^0_{n+m}
+{1\over {12}}n(n^2-1)\delta_{n+m,0}\quad .
\ee

  The eigenvalues of the Hamiltonian (4.15) are easily determined.
If the vacuum is associated with zero energy,
then by acting on $|0>$ with $\alpha^\pm_{-n}$ for $n>0$,
we obtain a state with energy equal to ${n\over{\kappa}}$.
By acting on $|0>$ with
$\beta_{-{n\over 2}}$ for $n>0$, we obtain a state with energy
 equal to ${n\over{2\kappa}}$.

 \sxn{ The Question of Conformal Symmetry}

Normally, the existence of Virasoro algebras indicates that the
system is conformally invariant.  However, the notion of conformal
invariance for fields defined on manifolds such as a circle and
on networks are quite different.
We will make this fact evident below.

The action (2.14) for fields on the figure eight for
arbitrary $\kappa_a$ and $\nu_a$ can be written in the form
\be
S=S_1+S_2\;,\qquad
S_a=\nu_a\kappa_a\int d\sigma^+_a d\sigma^-_a\;
{{\partial \chi_a }\over{\partial\sigma^+_a}}
{{\partial\chi_a }\over{\partial\sigma^-_a}} \;\;,
\qquad\sigma_a^\pm =\kappa_a\sigma\pm t\;\;,
\ee
which by itself displays the usual conformal symmetries
\be
\sigma_a^+\rightarrow\sigma_a^+ + F^+_a(\sigma_a^+)\;,
\ee
\be
\sigma_a^-\rightarrow\sigma_a^- + F^-_a(\sigma_a^-)\;.
\ee
However, once we impose the boundary conditions for the fields on a
figure eight, the symmetry transformations (5.2) and (5.3) will not
be independent.  For infinitesimal $F^+_a$'s and $F^-_a$'s, the
fields $\chi_a$ undergo the variations
\be
\delta\chi_a={1\over{2\nu_a\kappa_a}}\biggl[
\Bigl(F^+_a(\sigma_a^+) -F^-_a(\sigma_a^-)\Bigr)J_t^a
+\Bigl(F^+_a(\sigma_a^+) +F^-_a(\sigma_a^-)\Bigr)J_\sigma^a
 \biggr] \quad .
\ee
Consistency with the boundary conditions (2.11) means that
$\delta\chi_1(0,t)=\delta\chi_1(2\pi,t)=
\delta\chi_2(0,t)=\delta\chi_2(2\pi,t)$.
{}From the first term in brackets
and the conditions (2.24), we then get
\begin{eqnarray}
& &F^+_1(t) -F^-_1(-t)  =
F^+_1(2\pi\kappa_1+t) -F^-_1(2\pi\kappa_1-t)
\nonumber\\
&=&F^+_2(t) -F^-_2(-t) =
F^+_2(2\pi\kappa_2+t) -F^+_2(2\pi\kappa_2-t)\;.
\end{eqnarray}
{}From the second term in brackets, we get
\begin{eqnarray}
0&= &F^+_1(t) +F^-_1(-t)  =
F^+_1(2\pi\kappa_1+t) +F^-_1(2\pi\kappa_1-t)
\nonumber\\
&=&F^+_2(t) +F^-_2(-t) =
F^+_2(2\pi\kappa_2+t) +F^-_2(2\pi\kappa_2-t)\;\;\;.
\end{eqnarray}
Upon combining eqs. (5.5) and (5.6), we have
\be
F^+_1(t)   =F^+_2(t)=-F^-_1(-t) =-F^-_2(-t)\;\; \equiv \;\;F(t)
\ee
and
\be
F(t)=F(2\pi\kappa_1+t) =F(2\pi\kappa_2+t)
=-F(2\pi\kappa_1-t) =-F(2\pi\kappa_2-t) \;.
\ee

Eqs. (5.7) and (5.8) state that all $F_a^\pm$ are given by just
 one independent function $F$ which is odd in $t$ and
simultaneously $2\pi\kappa_1$ periodic and $2\pi\kappa_2$ periodic.
\it This, of course, is possible only when
$\kappa_1 \over\kappa_2$ is rational (if the trivial case where $F$ is the
zero function is excluded).  So a nontrivial analogue of
conformal symmetry exists only in this case.  \rm
We shall assume that $\kappa_1 \over\kappa_2$ is rational in the rest of this
Section.
We note also that unlike the analogous field theory on a
circle, there do not exist separate left and right conformal
transformations. As a result,
there do not exist two commuting sets of conformal generators on
the figure eight, as there do on the circle.

We note that the transformations (5.2) and (5.3),
along with the restrictions
(5.7) and (5.8), preserve Kirchhoff's law for the currents
$J_\sigma^a =   \nu_a\partial_\sigma \chi_a $.  This follows from
\begin{eqnarray}
\delta  J_\sigma^a|_0 & =&\bigl(\partial_t F(t)\bigr)\;J^a_\sigma |_0
+{1\over{\kappa_a}}F(t) \partial_\sigma J^a_t |_0 \quad ,
\nonumber\\
\delta  J_\sigma^a|_{2\pi}&  =&\bigl(\partial_t F(t)\bigr)
\;J^a_\sigma |_{2\pi}
+{1\over{\kappa_a}}F(t) \partial_\sigma J^a_t |_{2\pi} \quad
\nonumber
\end{eqnarray}
and (2.23).\\
\indent
What are the generators of the transformation (5.2) and (5.3)?
According to Noether's theorem, for infinitesimal variations
$\delta\sigma_a^\pm$
which are such that the induced variations (5.4) of $\chi_a$
leave the action (5.1) invariant, one has\cite{gol}
$$
\sum_{a=1,2}{\nu_a\kappa_a}
\int d\sigma^+_a d\sigma^-_a\;
\biggl\{\partial_{\sigma_a^+}\biggl(
{{\delta {\cal L}_a }\over{\delta ( \partial_{\sigma_a^+}\chi_a)}}
\delta\chi_a  -{\cal L}_a \;\delta\sigma_a^+ \biggr)
 + \partial_{\sigma_a^-}\biggl(
{{\delta {\cal L}_a }\over{\delta (\partial_{\sigma_a^-}\chi_a)}}
\delta\chi_a -{\cal L}_a \;\delta\sigma_a^- \biggr)  \;\biggr\} \;=\;0\;,
$$
where
$$
{\cal L}_a=\partial_{\sigma_a^+}\chi_a \; \partial_{\sigma_a^-}\chi_a\;,
$$
and
$\partial_{\sigma_a^\pm}= {{\partial  }\over{\partial\sigma^\pm_a}}$.
Upon substituting transformations (5.2-4), we have
$$
\sum_{a=1,2}{1\over{4\nu_a\kappa_a}}
\int d\sigma^+_a d\sigma^-_a\;
\biggl\{\partial_{\sigma_a^+}(F^-_aJ_-^a J_-^a)
+\partial_{\sigma_a^-}(F^+_aJ_+^a J_+^a)\biggr\} \;\;=\;\;0\quad.
$$
This result can be written as a current conservation law.  By
changing variables from
$(\sigma^+_a,\;\sigma^-_a)$ to $ (\sigma, t)$, we get
$$
\int d\sigma dt\; \sum_{a=1,2}\biggl\{
 \kappa_a\partial_t j_t^a - \partial_\sigma j_\sigma^a\biggr\}=0\;,
$$
or
\be
\sum_{a=1,2}\biggl\{
 \kappa_a\partial_t j_t^a - \partial_\sigma j_\sigma^a\biggr\}=0\;,
\ee
where the currents $ j_t^a $ and $ j_\sigma^a$ are given by
\begin{eqnarray}
j^a_t&=&{1\over{8\nu_a\kappa_a}}
(F^-_aJ^a_-J^a_--F^+_aJ^a_+J^a_+)\;,
\nonumber\\
j^a_\sigma &=&-{1\over{8\nu_a\kappa_a}}
(F^-_aJ^a_-J^a_-+F^+_aJ^a_+J^a_+)\;.
\end{eqnarray}

The conserved charge $q(F)$ associated with these currents
is a linear combination of
$\int_0^{2\pi} d\sigma \; j^1_t$ and $\int_0^{2\pi} d\sigma \; j^2_t$
and can be obtained by integrating the time component of the Noether
current in (5.10), the result being
\begin{eqnarray}
q(F)& =&\sum_{a=1,2} \kappa_a \int_0^{2\pi} d\sigma \; j^a_t
\nonumber\\
&=&-\sum_{a=1,2}  {1\over{8\nu_a}}\int_0^{2\pi} d\sigma \;
\biggl(F(-\sigma_a^-)J^a_-J^a_-+F(\sigma^+_a)J^a_+J^a_+\biggr)\;.
\end{eqnarray}

The conservation law (5.9), by itself, does not guarantee that the
conformal charges are conserved in time.  We have
\ba
\partial_t q(F)\;&=&\;\sum_{a=1,2} \kappa_a
\int_0^{2\pi} d\sigma \; \partial_t j^a_t \nonumber\\
 &=&\;\sum_{a=1,2} \int_0^{2\pi} d\sigma \; \partial_\sigma j^a_\sigma \\
&=&\;\sum_{a=1,2} j^a_\sigma \bigg|_0^{2\pi}\; \;,    \nonumber
\ea
from which it follows that, in order for $q(F)$ to be constant in
time, the space component of the conformal current
has to fulfill Kirchhoff's law at the junction.
It can be checked that, thanks to the conditions (5.7) and
(5.8) on $F$ and the boundary conditions (2.22) and (2.24) on
$J_\sigma^a$ and $ J_t^a $, this is indeed the case:
\begin{eqnarray}
\sum_{a=1,2} j^a_\sigma \bigg|_0^{2\pi}
&=&-\sum_{a=1,2}{1\over{8\nu_a\kappa_a}}
(F^-_aJ^a_-J^a_-+F^+_aJ^a_+J^a_+)\bigg|_0^{2\pi}
\nonumber\\
&=&-F(t)\sum_{a=1,2}{1\over{8\nu_a\kappa_a}}
(J^a_+J^a_+-J^a_-J^a_-) \bigg|_0^{2\pi} \\
&=&-\frac{1}{2} F(t)\sum_{a=1,2}
{1\over {\nu_a\kappa_a}} J_\sigma^a J_t^a \bigg|_0^{2\pi} \;\;
\nonumber \\
&=& \;\;0\;.
\nonumber
\end{eqnarray}

One can show that $q(F)$
is differentiable with respect to variations in $\chi_a(\sigma)$ and
the canonical momenta $\kappa_a J^a_t$.  That is, that no boundary terms
appear in the resulting variations of $q(F)$.  Of course, this is obvious
for variations in $ J^a_t$.
Concerning variations in $\chi_a$, upon substituting
$\delta J_\pm^a=\nu_a\partial_\sigma \delta\chi_a$ into (5.11) we
obtain the boundary term
$$
-{1\over 4}\sum_{a=1,2}
\biggl(F(-\sigma_a^-)J^a_-+F(\sigma^+_a)J^a_+\biggr)\delta \chi_a
\bigg|_{\sigma=0}^{\sigma=2\pi}  \;.
$$
However,
if we again assume continuity of the phase at the junction, so that
$ \delta\chi_1(0,t)=$   $\delta\chi_1(2\pi,t)
=$      $\delta\chi_2(0,t)=$        $\delta\chi_2(2\pi,t)$,
along with the result (5.8), this boundary term reduces to
$$
-{1\over 2}\sum_{a=1,2}
F(t)(J^a_-+J^a_+)\bigg|_{\sigma=0}^{\sigma=2\pi}   \delta \chi_a   \;,
$$
which then vanishes by Kirchhoff's law (2.22).

The variational derivatives of $q(F)$ with respect to
$ J^a_t$ and $\chi_a$ are given by
\ba
{{\delta q(F)}\over{\delta J^a_t(\sigma)}}&=&
{1\over{4\nu_a}}\biggl(F(-\sigma_a^-)J_-^a -
F(\sigma_a^+)J_+^a  \biggr)    \quad{\rm and}\quad \nonumber\\
{{\delta q(F)}\over{\delta \chi_a(\sigma)}}&=&
{1\over 4}\partial_\sigma\biggl(
F(-\sigma_a^-)J_-^a + F(\sigma_a^+)J_+^a  \biggr)
\ea

Using eqs. (3.10) and (5.14), we can compute the Poisson brackets
between a smeared current ${\cal J}(\Lambda)$ and the conformal
charge $q(F)$:
$$
\{{\cal J}(\Lambda),q(F)\}=
\sum_{a=1,2} {1\over{2\kappa_a}}\int_0^{2\pi} d\sigma \;
\biggl( F(\sigma_a^+)\partial_\sigma\Lambda_+^a J^a_+ -
       F(-\sigma_a^-)\partial_\sigma\Lambda_-^a J^a_- \biggr)
$$
\be
\quad \qquad -\; F(t)
\sum_{a=1,2}{1\over{4\kappa_a}}
(\Lambda^a_+ -\Lambda^a_-)(J_+^a+J^a_-)
\bigg|^{\sigma=2\pi}_{\sigma=0}   \quad.
\ee

For field theory on a circle, the Poisson bracket
between a conformal generator and a current is still a current.
This result does not seem to generalize to the figure eight.
This is so firstly because of the boundary term in eq. (5.15).  Further,
the integral in eq. (5.15) cannot in general be replaced by
a smeared current ${\cal J}(\bar \Lambda)$ for a test function
$\bar\Lambda\in {\cal T}$.  This is because what stands for
$\bar\Lambda$ in the integral (5.15)
does not satisfy the condition (3.9)
and hence does not belong to the test function space ${\cal T}$.

The Poisson brackets between two conformal charges $q(F)$ and $q(\bar F)$
defines the classical Virasoro or the Witt algebra.  We get
$$
\{q(F),q(\bar F)\}=q(\bar{\bar F})     \quad,
$$
where
\be
\bar{\bar F}(\sigma) \equiv {1\over 2}(\bar F \partial_\sigma F
- F \partial_\sigma \bar F)(\sigma)   \;,
\ee
and we have used the conditions (5.7) and (5.8)
to eliminate boundary terms.
Eq. (5.16) is the standard relation defining the Witt algebra.

Let us rewrite (5.16) in terms of Fourier modes.  Let
the smallest period of
the periodic function $F$ be $2\pi\kappa$.
Then in view of (5.8),
$\kappa_a$ has to be an integer multiple of $\kappa$:
 $$\kappa_a=N_a\;\kappa\;,\quad N_a={\rm integer}\;.$$
We now define the Fourier components $L_n$ of the conformal charge
 as follows:
\be
L_n=-2\kappa\; q(e^{in\sigma/\kappa})  \;\;.
\ee
The Poisson bracket of $L_n$ with $L_m$ is then a familiar one:
\be
\{L_n\;,\;L_m\}=-i(n-m)L_{n+m} \;.
\ee

 If we now specialize to the case ${\bf b}$ where $\kappa_1=\kappa_2
 =\kappa$ and $\nu_1=\nu_2=\nu$, and apply the expansions
 (4.8) and (4.9), then $L_n$ can be written as
\begin{eqnarray}
L_n&=&{\kappa\over{8\nu}}\;e^{int/\kappa}
\int_{0}^{2\pi}d\sigma \;\biggl\{
e^{in\sigma}\biggl((J^{sum}_+)^2 +(J^{dif}_+)^2 \biggr)
+e^{-in\sigma}\biggl((J^{sum}_-)^2 +(J^{dif}_-)^2 \biggr) \biggr\}
\nonumber\\
&=&e^{int/\kappa}\biggl(L^+_n+L^-_n+{1\over 2}L^0_{2n}\biggr) \quad,
\end{eqnarray}
where $L^+_n$, $L^-_n$ and $L^0_{n}$ were defined in eqs. (4.13).
$L_n$ is thus the sum of three Virasoro generators
which commute in quantum theory.  In view of (4.15),
we further obtain the result that the zero component $L_0$ of the
algebra is the generator of time translations,
that is that it is proportional to the Hamiltonian
${\cal H}={1\over{\kappa}}L_0$.

It is easy to verify (5.18) starting from the Poisson brackets (4.14).

So far our treatment of the figure eight has been purely classical.
In quantum theory, we pick up an additional anomaly term in the
Witt algebra defined by (5.18).  If we regularize the theory
so that the central terms for the algebra generated by
$L^+_n$, $L^-_n$ and $L^0_{n}$ in quantum theory
have the standard form as in eqs. (4.16),
then the central term in the commutator $[ L_n\;,\;L_m ] $  will be
\be
{1\over {24}} n(8n^2-5)\delta_{n+m,0}
\quad.
 \ee

\sxn {Absence of Chiral Currents in Quantum Theory}

Here we show that the chiral currents
$J_+^a(\sigma)$ and $J_-^a(\sigma)$
cannot be {\it independently} quantized on the figure eight.
More precisely, the two chiral currents cannot be expanded
in terms of two independent sets of bases, which
i) when quantized have a well defined action on the Fock space,
and ii) lead to the correct Poisson brackets between the
chiral currents.
  We can state this claim in another way.  Let us define
left- and right-moving smeared classical currents, which we denote by
${\cal J}_+(\Lambda)$ and
${\cal J}_-(\Lambda)$ respectively, according to
\be
{\cal J}_+(\Lambda)=\sum_{a=1,2}\int d\sigma\;
\Lambda_+^a (\sigma)J_+^a(\sigma)
\quad {\rm and} \quad
{\cal J}_-(\Lambda)=\sum_{a=1,2}\int d\sigma\;
\Lambda_-^a (\sigma)J_-^a(\sigma)
\qquad .
\ee
Then there do not exist two subsets ${\cal T_+}\in {\cal T}$ and
${\cal T_-}\in {\cal T}$
of test functions of the form
$$
\Lambda^{(A^+_n)}=   \; (f^1_n,\; f^2_n,\;0,\; 0)
$$
\be
\Lambda^{(A^-_n)}=   \; (0,\;0,\;g^1_n,\; g^2_n)
\ee
satisfying the properties of orthonormality and completeness,
$$
\sum_{a=1,2} \int_0^{2\pi} d\sigma \; f_n^{a *}(\sigma) f^a_m(\sigma)=
\delta_{n,m}   \quad ,
$$
\be
\sum_n \; f_n^{a}(\sigma) f^{a *}_n(\sigma ^{\prime})=
\delta ^a (\sigma - \sigma ^{\prime})   \quad ,
\ee
and
$$
\sum_{a=1,2} \int_0^{2\pi} d\sigma \; g_n^{a *}(\sigma) g^a_m(\sigma)=
\delta_{n,m}   \quad ,
$$
\be
\sum_n \; g_n^{a}(\sigma) g^{a *}_n(\sigma ^{\prime})=
\delta ^a (\sigma - \sigma ^{\prime})   \quad ,
\ee
such that
i) the quantum operators $A^+_n$ and $A^-_n$ corresponding respectively to
${\cal J}_+(\Lambda^{(A^+_n)})$ and ${\cal J}_-(\Lambda^{(A^-_n)})$
have a well defined action on the Fock space, and
ii) the chiral currents
\be
\hat {J}_+^a(\sigma) =\sum_n  A^+_n f^a_n (\sigma)  \;\;,~~
\hat {J}_-^a(\sigma) =\sum_n  A^-_n g^a_n (\sigma)
\ee
give the correct Poisson brackets, eq. (3.12), for the corresponding
classical observables. [Here $\delta^a$ denotes the $\delta$ function
corresponding to loop $a$.]\\
\indent
 For simplicity, we shall prove the result for case ${\bf b}$
defined by $\kappa_1=\kappa_2=\kappa$ and $\nu_1=\nu_2=\nu$.
The proof can easily be generalized to any case.
The mode expansions for the currents in case ${\bf b}$ are
given in eqs. (4.8) and (4.9).  (The mode expansions for the
fields appear in Appendix B.)

The proof is by contradiction.  We suppose that two complete and
orthonormal sets of test functions, $\{\Lambda^{(A^+_n)}\}$
and $\{\Lambda^{(A^-_n)}\}$
satisfying the above conditions
exist. Then, the condition of completeness implies that
$$
J_+^a(\sigma) =\sum_n  {\cal J}_+(\Lambda^{(A^+_n)})
  \;f^a_n (\sigma)  \quad ,
$$
\be
J_-^a(\sigma) =\sum_n  {\cal J}_-(\Lambda^{(A^-_n)})
  \;g^a_n (\sigma)  \quad ,
\ee
for any classical currents $J_+^a(\sigma)$ and $J_-^a(\sigma)$.
In quantum theory, we let $A_n^+$ ($A_n^-$) be the operators corresponding
to $ {\cal J}_+(\Lambda^{(A^+_n)}) $ ($ {\cal J}_-(\Lambda^{(A^-_n)}) $).
Further, let $|0>$ be the vacuum state in the Fock space on which
the quantum operators $A^+_n$ and $A^-_n$ can act.
We first show that in order for $A^+_n |0>$ to have finite norm,
$f^a_n$ must be continuous at the junction.
Moreover, the limiting value of $f^1_n$ at the junction must be
 the same as that of $f^2_n$.
Analogous results apply to the functions $g^a_n$.\\
\indent
To proceed let us recall that the test function space
$ {\cal T}$ for the case $\kappa_1=\kappa_2=\kappa$ and
$\nu_1=\nu_2=\nu$ is spanned by $\Lambda^{(\alpha^\pm_n)} $
and $\Lambda^{(\beta_{n\over 2})}$ defined in eqs. (4.10).
Therefore the left moving current
$\tilde{J}^a_+$ has the expansion
\be
\tilde{J}_+^a(\sigma)~=~-{1\over 2}\sum_m \alpha_m^+ e^{-im\sigma}
-{{(-1)^a}\over 2}\sum_m \beta_{m\over 2}\; e^{-im\sigma /2}   \quad,
\ee
$\alpha^\pm_m$ and $\beta_{m\over 2}$ now being quantum operators.
[For economy of notation, we will not introduce symbols for them
distinct from those in (4.8) and (4.9).]
Substituting it into the expression for
${\cal J}_+(\Lambda^{(A^+_n)})$, we obtain an expression for
$A^+_n$ in terms of $\alpha^+_n $ and $\beta_{n\over 2}$:
\be
A^+_n=\sum_m \alpha^+_m N_{n,m} +\sum_m \beta_{m\over 2} M_{n,m}\quad,
\ee
where
\be
N_{n,m}=-{1\over 2}\int d\sigma \; (f^1_n +f^2_n)^{*} e^{-im\sigma}
\quad {\rm and} \quad
M_{n,m}=-{1\over 2}\int d\sigma \; (f^2_n -f^1_n)^{*}
 e^{-im\sigma /2} \quad .
\ee
If we now apply the quantum analogues
\be
[\alpha^+_{m},\;\alpha^+_{-n}
]= {{2 n \nu }\over {\pi\kappa}}   \;\delta_{m,n}     \;,\quad
[\beta_{m \over 2},\;\beta_{-{n \over 2}}
]={{n\nu }\over {2\pi\kappa}}   \;\delta_{m,n}  \;,
 \ee
 of the Poisson brackets relations (4.12) and assume
that $\alpha^+_n$ and $\beta_{n\over 2}$ annihilate
the vacuum when $n>0$, we obtain the following expression for the
squared norm of the state  $A^+_n |0>$:
\be
\Big|A^+_n |0>\Big|^2 =
\Big| (N_{0,0}\; \alpha^+_0 +M_{0,0} \;\beta_0 )|0>\Big|^2
+{{2\nu}\over{\pi\kappa}}
\sum_{m>0} \Bigl(m|N_{n,m}|^2 + {m\over 4} |M_{n,m}|^2\Bigr)
\quad.  \ee

By integrating by parts twice, we can rewrite $N_{n,m}$ and
$M_{n,m}$ according to
\begin{eqnarray}
N_{n,m}&=&-{i\over{2m}}(f^1_n+f^2_n)^{*}|^{2\pi}_0   +
O({1\over {m^2}})\;, \nonumber\\
M_{n,m}&=&
-(-1)^m{i\over{m}}(f^2_n-f^1_n)^{*}|_{2\pi}
+{i\over{m}}(f^2_n-f^1_n)^{*}|_0  +
O({1\over {m^2}})\;. \nonumber
\end{eqnarray}
Substituting the above into eq. (6.11), we have
\begin{eqnarray}
\Big|A^+_n |0>\Big|^2
&=&  \Big| (N_{0,0}\; \alpha^+_0 +M_{0,0} \;\beta_0 )|0>\Big|^2
\nonumber\\
& &    +{{\nu}\over{\pi\kappa}}      \;\;
\sum_{m\;{\rm even}, > 0}\;\; {1\over m}\biggl(
\Big|(f^1_n|^{2\pi}_0) \Big|^2 + \Big|(f^2_n|^{2\pi}_0) \Big|^2 +
 O({1\over {m}})\biggr)   \\
& & +{{\nu}\over{2 \pi\kappa}}
\sum_{m\;{\rm odd}, > 0} {1\over m}\biggl(
\Big|(f^1_n - f^2_n)_{2\pi} + (f^1_n - f^2_n)_0 \Big|^2 +
O({1\over {m}})\biggr)  \;.  \nonumber
\end{eqnarray}
In order for the first summation in (6.12) to be
convergent, we must require that
$ f^1_n|^{2\pi}_0 = f^2_n|^{2\pi}_0 =0 $,
while in order for the second summation also to be
convergent, we must in addition have
$f^1_n|_0 = f^2_n|_0 =0 $.
Thus the functions $f^a_n$ for $a=1,2$ must
have a unique value at the junction:
\be
 f^1_n|_{2\pi} = f^2_n|_{2\pi} = f^1_n|_{0} = f^2_n|_{0} \quad.
\ee

The same argument can be applied to the test functions
$\Lambda^{(A^-_n)}$ of the right moving currents, from which one
finds that the functions $g^a_n$, for $a=1,2$ must
have a unique value at the junction:
\be
 g^1_n|_{2\pi} = g^2_n|_{2\pi} = g^1_n|_{0} = g^2_n|_{0} \quad.
\ee

An immediate consequence of eqs. (6.13) and (6.14) is that the
 Poisson brackets of ${\cal J}_+(\Lambda^{(A_n^+)})$ and
${\cal J}_-(\Lambda^{(A_m^-)})$, and hence the corresponding
commutators between $A^+_n$ and $A^-_m$, vanish.
In fact from eq. (3.12) it follows that
$$
\{{\cal J}_+(\Lambda^{(A_n^+)}),{\cal J}_-(\Lambda^{(A_m^-)})\}\;=\;
\{{\cal J}(\Lambda^{(A_n^+)}),{\cal J}(\Lambda^{(A_m^-)})\}\;$$
\be
=\;{{\nu}\over{\kappa}}\sum_{a=1,2}
f^{a *}_n  g^{a *}_m \bigg|^{\sigma=2\pi}_{\sigma=0}\;=\;0 \quad.
\ee
 From this and eqs. (6.5), we must also then conclude that the
commutator between $\tilde{J}_+^a(\sigma)$ and $\tilde{J}_-^a(\sigma')$
vanishes.

But from eqs. (4.3), (4.8) and (4.9)
we also have, for $0 \leq \sigma \leq 2 \pi$,
$$J^1_+(\sigma)~=~\frac{1}{2}(J^{sum}_+(\sigma)-J^{dif}_+(\sigma))~=
\frac{1}{2}(J^{sum}_+(\sigma)-K(\sigma))~$$
\be
=-\frac{1}{2} \sum_n \alpha_n^+ e^{-in\sigma} + \frac{1}{2}
\sum_n \beta_{\frac{n}{2}} e^{-i\frac{n}{2}\sigma}
\ee
and
$$J^1_-(\sigma)~=~\frac{1}{2}(J^{sum}_-(\sigma)-J^{dif}_-(\sigma))~=
\frac{1}{2}(J^{sum}_-(\sigma)-K(4 \pi - \sigma))~$$
\be
=\frac{1}{2} \sum_n \alpha_n^- e^{in\sigma} + \frac{1}{2}
\sum_n \beta_{\frac{n}{2}} e^{i\frac{n}{2}\sigma}  \;.
\ee
It follows that
$$\{ J^1_+(\sigma),J^1_-(\sigma^{\prime}) \}= \frac{1}{4}
\sum_{n,m} \{ \beta_{\frac{n}{2}}, \beta_{\frac{m}{2}} \}
e^{-i\frac{n}{2}\sigma}e^{i\frac{m}{2}\sigma^{\prime}}~$$
\be
=-\frac{1}{4} \sum_n \frac{in \nu}{2 \pi k} e^{-i \frac{n}{2}(\sigma +
\sigma^{\prime})}~\neq~0  \;\;.
\ee
A similar result holds on loop 2.\\
\indent
We thus see that if we try to
quantise the chiral components of the currents separately, we
get wrong commutation relations for them. This completes the
proof.

\bigskip

{\bf Acknowledgements}

We have been supported during the course of this work as follows: 1)
A. P. B. and G. B. by the Department of Energy, USA,
under contract number
DE-FG-02-85ER-40231, 2) K. G. by the Department of Energy, USA,
under contract number DE-FG02-91ER40685,
3) A. Stern by the Department of Energy, USA under
contract number DE-FG-05-84ER-40141,
 4) G. B., G. M. and A. Simoni by the Dipartimento
di Scienze Fisiche, Universit{\`a} di Napoli, and 5) P. S. by the Swedish
National Science Research Council under contract number F-FU 8230-303.
P. S. and A. Stern wish to thank the
group in Naples, Giuseppe Marmo, in particular,
for their hospitality while this work was in progress.
We are grateful for discussions with F. Lizzi,
S. Rajeev, D. Sen and F. Zaccaria.
We thank T. Pennington for the figure.


\bigskip
\centerline {{\bf APPENDICES}}

\axn{Self-Adjoint Extensions}

Here we examine boundary conditions more general than the ones
we specified in Section 2 [cf. eq. (2.11)
]. As we stated there, the boundary conditions must be so chosen
that the operator $H$, defined formally in eq. (2.7), is
self-adjoint.  For this purpose,
to start with, we can choose a domain ${\cal D}_0$ such that the
restriction $H_0$ of $H$ to it is symmetric.  This means by definition that
\be
<\tilde\psi^0,\;H_0 \tilde\chi^0> =<H_0 \tilde\psi^0 ,\;\tilde\chi^0>\;,
{}~~~~~\forall\; \tilde\chi^0, \tilde\psi^0 \in {\cal D}_0     \;,
\ee
where the scalar product was defined in eq. (2.8).
This equation is equivalent to
\begin{eqnarray}
0&=&\sum_{a=1,2} \nu_a\int d\sigma\;
\tilde\psi^{0*}_{ a}(\sigma) \partial_\sigma^2 \tilde\chi^{0}_{a}(\sigma)
 -\sum_{a=1,2} \nu_a \int d\sigma\;
\partial_\sigma^2 \tilde\psi^{0*}_{a}(\sigma)
 \tilde\chi^{0}_{a}(\sigma)
\nonumber\\
&=&-\sum_{a=1,2} \nu_a \biggl(\partial_{\sigma}
\tilde \psi^{0*}_{a} \tilde \chi^{0}_{a} -
 \tilde \psi^{0*}_{a}
 \partial_{\sigma} \tilde \chi^{0}_{a}\biggl)\bigg|_0^{2 \pi} \quad.
\end{eqnarray}
This condition is certainly fulfilled if ${\cal D}_0$ is taken to be
the set of functions which vanish at the junction together with
their first derivatives:
\begin{eqnarray}
{\cal D}_0 \equiv \{\tilde \chi ^0\;|\;
 \tilde\chi^0_1(0)&=&\tilde\chi^0_2(0)=\tilde\chi^0_1(2 \pi)=
\tilde\chi^0_2(2 \pi)
\nonumber\\
&=&\partial_{\sigma}\tilde\chi^0_1(0)=
\partial_{\sigma}\tilde\chi^0_2(0)=
\partial_{\sigma}\tilde\chi^0_1(2 \pi)=
\partial_{\sigma}\tilde\chi^0_2(2 \pi) =0\;\;\}\;.
\nonumber
\end{eqnarray}

The operator $H_0$ is not self-adjoint in view
of the remark preceding eq. (2.13) since we can
 check that the domain ${\cal D}_0^{\dagger}$ of its adjoint,
 $H_0^{\dagger}$, is larger than ${\cal D}_0$.  We recall
that according to eq. (2.12),
 ${\cal D}_0^{\dagger}$  is defined
to be the set of all functions $\overline {\psi}$ fulfilling
\be
<\overline {\psi},\;H_0 \tilde\chi ^0>
=<H_0^{\dagger} \overline {\psi},\;\tilde\chi ^0>\;,  ~~~~~~\forall
\tilde\chi^0 \in {\cal D}_0        \;.
\ee
This is equivalent to
\begin{eqnarray}
0&=&\sum_{a=1,2} \nu_a\int d\sigma\;
\overline {\psi}_a^*(\sigma) \partial_\sigma^2
\tilde\chi^0_{a}(\sigma)
 -\sum_{a=1,2} \nu_a\int d\sigma\;
\partial_\sigma^2 \overline {\psi}_{a}^*(\sigma)
\tilde\chi^0_{a}(\sigma)
\nonumber\\
&=&-\sum_{a=1,2} \nu_a
 (\partial_{\sigma} \overline {\psi}_{a}^* \tilde\chi^0_{a} -
 \overline {\psi}_{a}^*
  \partial_{\sigma} \tilde\chi^0_{a})\bigg|_0^{2 \pi}   \quad .
\end{eqnarray}
In order to satisfy this equation, neither $\overline {\psi}$, nor its
derivatives need vanish at the junction.

This means that, in order to
make $H_0$ self-adjoint, we have to extend it to a domain larger
than ${{\cal D}_0}$.  Whether this can be done and in how many ways, is
determined by the deficiency index theorem,
which we now briefly review \cite{Sim}.

The deficiency indices ${\cal N}_+$ and ${\cal N}_-$
of $H_0$ are defined to be the
number of linearly independent orthonormal eigenvectors
$\overline {\psi}_m ^{(+)}~ [~m=1,\dots {\cal N}_+]$ and
$\overline {\psi}_n ^{(-)}~ [~n=1, \dots {\cal N}_-]$
of $H_0^{\dagger}$ in ${\cal D}_0^{\dagger}$ with
eigenvalues $+i$ and  $-i$ respectively:
$$
(H_0^{\dagger}~ \overline {\psi}_m ^{(+)})_a=
-{1\over{\kappa_a^2}}~\partial_\sigma^2~
\overline {\psi}_{ma}^{(+)}(\sigma)=
i~\overline {\psi}_{ma}^{(+)}(\sigma)~,~~~m=1, \dots {\cal N}_+\;,\quad
\overline {\psi}_{m}^{(+)} \in   {\cal D}_0^{\dagger}\;,
$$
$$
(H_0^{\dagger} ~\overline {\psi}_n ^{(-)})_a=
-{1\over{\kappa_a^2}}~\partial_\sigma^2~
\overline {\psi}_{na}^{(-)}(\sigma)=
-i~\overline {\psi}_{na}^{(-)}(\sigma)~,~~~n=1, \dots {\cal N}_-,\;\quad
\overline {\psi}_{m}^{(-)} \in   {\cal D}_0^{\dagger}\;,
$$
\be
< \overline {\psi}_{m}^{(\epsilon)}\;,\;,
 \overline {\psi}_{n}^{(\epsilon ')}> = \delta_{m,n}~
\delta_{\epsilon ,\epsilon '}  \;,\quad
\epsilon ,\epsilon '=\pm\;.
\ee
According to
the deficiency index theorem,  $H_0$ admits self-adjoint
extensions if and only if ${\cal N}_+ ={\cal N}_-= {\cal N}$.
With ${\cal N}_+ ={\cal N}_- ={\cal N}$, the
self-adjoint extensions of $H$ are in one-to-one
correspondence with $U({\cal N})$
matrices $\{g\}$.  Their domains ${\cal D}_g$
are direct sums of ${\cal D}_0$ with the vector space
spanned by the vectors $\psi _i = \overline {\psi}_i ^{(+)} + g_{i j}
\overline {\psi}_j^{(-)} ,\;1\le i\le {\cal N}$:
\be
{\cal D}_g={\cal D}_0 \oplus \{ {\rm span}~
( \overline {\psi}_i ^{(+)} + g_{i j}
\overline {\psi}_j^{(-)}) \}~,~g \in U({\cal N})~.
\ee

It is easy to check that both of the deficiency indices of
$H_0$ are equal to
$4$, which implies the existence of a sixteen-fold
infinity of self-adjoint
extensions. It can be shown that the domains corresponding to any given
choice of a matrix in $U(4)$ can also be described in terms of
boundary conditions involving
the functions and their first derivatives at the junction.  Functions
fulfilling a particular choice of these boundary conditions form a domain
${\cal D}^{(h)}$, $h \in U(4)$. It can be shown that ${\cal D}^{(h)}
{}~=~{\cal D}_g$
for some $g$.
These boundary
conditions are such that if the surface term in eq. (A.4) is to
vanish for
all functions $\chi_a(\psi_a)$ in ${\cal D}^{(h)}$, then
$\overline {\psi}_a(\chi_a)$ as well has to belong to ${\cal D}^{(h)}$.

The domain ${\cal D}(H)$ of (2.11) is
 ${\cal D}^{(h)}$ for a particular choice of $h$.
\\

\axn{ Mode Expansion }

Here we shall examine the general solutions of the field equations on a
figure eight consistent with the boundary conditions (2.10), (2.22) and
(2.24), and carry out the eigenmode expansions
for two special choices of the parameters $\kappa_a$ and $\nu_a$, namely:
\hfill\break
{\bf a. } $\;\kappa_1\nu_1=\kappa_2\nu_2\;$ and
{\bf b. } $\;\kappa_1=\kappa_2=\kappa\;$ and $\;\nu_1=\nu_2=\nu\;$.
For case {\bf a }, unlike in earlier Sections, we will in addition assume
that
$\;{\kappa_1}\over {\kappa_2}$ is irrational for reasons of simplicity.
Our aim is to find the basis
of test functions $\Lambda$ for the currents
${\cal J}(\Lambda)$ for the two cases.  For case
 ${\bf b }$, we show that our answer agrees with eq. (4.10).

The discussion which now follows is general and does not assume case
{\bf a } or {\bf b } until it is otherwise stated.

We first expand $\chi_a(\sigma,t)$ according to
\be
\chi_a(\sigma,t)=q + pt + N_a \sigma +
\sum_n \chi_a^n(\sigma)e^{i\omega_n t}
\;.
\ee
$q$ and $p$ are constants corresponding to zero frequency modes, while
$\chi_a^n(\sigma)$ denote the oscillatory modes.
The latter satisfy the equations
\be
\Bigl [ H_a -\omega_n^2 \Bigr ]   \;\chi_a^n(\sigma) = 0 \;,\qquad
H_a=- {1\over{\kappa_a^2}}\partial_\sigma^2 \;.
\ee
As in Section 2, we shall assume
that $\chi_a^n$ are singlevalued at the junction, so that
$\chi_1^n(0)=\chi_1^n(2\pi)=$    $\chi_2^n(0)=\chi_2^n(2\pi)$.
Since the phases
$\chi_a(0,t)$ and $\chi_a(2\pi,t)$ can differ only by $2\pi\; \times $
integer, the constants $N_a$ must take on integer values.
$N_a$ parametrize the ``winding modes".

For the solutions of eq. (B.2), we can take
$\chi_a^n(\sigma)= A_{a,n} \cos{k_{a,n}\sigma} +
B_{a,n} \sin{k_{a,n}\sigma}$  where $k_{a,n}=\kappa_a\omega_n\; (>0)$
if $\chi_a^n(\sigma)\ne 0$, and
the coefficients $A_{a,n}$ and $B_{a,n}$ are determined from
the boundary conditions.  [The value of $k_{a,n}$ is immaterial if
 $\chi_a^n(\sigma) = 0$.  Also the case
 $k_{a,n}=-\kappa_a\omega_n\;(<0)$ need not be separately considered
as it can be brought back to the present form by letting
$B_{a,n} \rightarrow -B_{a,n} $.  ]
For the latter, from the singlevaluedness conditions, we get
\be
A_{1,n}\;
=\;A_{2,n}\;=\;A_{1,n} \cos{2\pi k_{1,n}} +B_{1,n} \sin{2\pi k_{1,n}}
\;=\; A_{2,n} \cos{2\pi k_{2,n}} +B_{2,n} \sin{2\pi k_{2,n}}\;.
\ee
In addition, the Kirchhoff law (2.22) gives
\be
\sum_{a=1,2}\nu_a \kappa_a \biggl( A_{a,n} \sin{2\pi k_{a,n}} +
B_{a,n} (1-\cos{2\pi k_{a,n}})\biggr)=0\;.
\ee

Eqs. (B.3) and (B.4) form a system of homogeneous linear equations
 for $A_{a,n}$ and $B_{a,n}$.
Solutions for $A_{a,n}$ and $B_{a,n}$ exist provided
the determinant of the associated matrix is zero, that is,
\be
 \nu_1 \kappa_1 (1-\cos {2\pi k_{1,n}})\sin{2\pi k_{2,n}} +
\nu_2 \kappa_2  (1-\cos {2\pi k_{2,n}})\sin{2\pi k_{1,n}}=0\;.
\ee
Using this equation, we can classify five types of solutions for
 $k_{a,n}$, along with their corresponding eigenmodes
($\chi_1^n$, $\chi_1^n$).  They are:
\bigskip

i) $k_{1,n}=n$ is a positive integer, and
 ($\chi_1^n(\sigma)$, $\chi_1^n(\sigma)$)=($\sin{n\sigma}$, 0).

\bigskip
ii)  $k_{2,n}=n$ is a positive integer, and
($\chi_1^n(\sigma)$, $\chi_1^n(\sigma)$)=(0, $\sin{n\sigma}$).

As $\chi_2^n$ ($\chi_1^n$) is zero
in case i) ( ii) ), the value  of $k_{2,n}$ ( $k_{1,n}$ )
in that case  is immaterial.

\bigskip
iii)
If ${{\kappa_1} \over {\kappa_2}}$ is rational, we also have the
solutions $k_{a,n}=n_a$ = integer

where ${{n_{1}}\over {n_{2}}}={{\kappa_1} \over {\kappa_2}}$, and
($\chi_1^n(\sigma)$, $\chi_1^n(\sigma)$)=($\cos{n_1\sigma}$,
$\cos{n_2\sigma}$).

\bigskip
iv) Both $2 k_{1,n}=2 r_1$ and $2 k_{2,n}= 2 r_2$ are positive odd
integers and

($\chi_1^n(\sigma)$, $\chi_1^n(\sigma)$)=
($\nu_2 \kappa_2 \sin{r_1\sigma}$, $-\nu_1 \kappa_1  \sin{r_2\sigma}$).
Just as for iii), these modes are

possible only when ${{\kappa_1} \over {\kappa_2}}$ is rational.

\bigskip
v) Neither $2 k_{1,n}$ nor $2 k_{2,n}$ are integers.
Rather, $k_{a,n}$ are positive solutions

 of the transcendental equation
\be
 \nu_1 \kappa_1 \tan{\pi k_{1,n}} +
 \nu_2 \kappa_2 \tan{\pi k_{2,n}}=0\;.
\ee

Then the corresponding eigenmodes are given by
\be
\chi_a^n(\sigma)=
\cos{k_{a,n}\sigma} +  \tan{\pi k_{a,n}} \sin{k_{a,n}\sigma}\;.
\ee

\bigskip
Solutions i) and ii) correspond to independent oscillations on loops
1 and 2 respectively, and exist for arbitrary values of the
independent parameters $\kappa_a$ and $\nu_a$.  On the other hand,
the presence of modes iii-v) depends on the values of
$\kappa_a$ and $\nu_a$.

The solutions can
be used to form an orthonormal basis with respect to the scalar product
(2.8).  If $\chi$ and $\psi$ are eigenmodes for distinct eigenvalues,
so that $\Bigl[ H_a -\omega^2 \Bigr]
\;\chi_a(\sigma) =  \Bigl[  H_a -{\omega '}^2 \Bigr ]
\;\psi_a(\sigma) = 0 $ and $\omega^2\not =\omega '^2$, then
 $\chi$ and $\psi$ are of course orthogonal with respect to the scalar
product.  Orthogonal combinations of degenerate eigenmodes can be formed,
and the modes can be normalized as well.
 The completeness of the eigenvectors $(\chi^n_1,\;\chi^n_2)$
follows from the result that the operator
$H=(H_1,\;H_2)$ is self-adjoint, as was shown in Section 2.

We now expand the fields $\chi_a$ in terms of the eigenmodes i-v)
for cases ${\bf a}$ and ${\bf b}$.
As case ${\bf b}$ is the simpler of the two, we begin with it.

\centerline{
{\bf b.} $\kappa_1=\kappa_2=\kappa$ and $\nu_1=\nu_2=\nu$ }

The quantities $\ell_a$, $T_a$ and $v_a$ of Sections 1 and 2 correspond
to length, tension and velocity respectively on loop $a$.  This case
requires that the ratio of these quantities for the
two loops must be the same:
${{\ell_1}\over{\ell_2}}={{T_1}\over{T_2}}={{v_1}\over{v_2}}\;.$
  In particular, if $T_a$ and $v_a$ are the same for both
loops, then so must be the lengths $\ell_a$.

For this case, the solutions for $2k_{a,n}$ can only be integers
as follows from (B.5),$\nu_a\kappa_a$ and $k_{a,n}=\kappa_a\omega_n$
being independent of $a$.  Therefore
type v) eigenmodes are not present in the expansion
for $\chi_a(\sigma,t)$.  The expansion can be written as
\be
 \chi_1(\sigma,t)=Q( t) + N_1\sigma
 +\sum_{m=1}^\infty \biggl\{a_{1,m}(t)\;\sqrt{2}\sin{m\sigma}
+b_{m}(t)\;\cos{m\sigma}\biggr\}
+\sum_{r={1\over 2}}^\infty  c_{r}(t)\;\sin{r\sigma}
\ee
and
\be
 \chi_2(\sigma,t)=Q( t) + N_2\sigma
 +\sum_{m=1}^\infty \biggl\{a_{2,m}(t)\;\sqrt{2}\sin{m\sigma}
+b_{m}(t)\;\cos{m\sigma}\biggr\}
-\sum_{r={1\over 2}}^\infty  c_{r}(t)\;\sin{r\sigma} \;,
\ee
where $m$ is a positive integer and $2r$ is a positive odd integer
[so that $r={1\over 2} ,{3\over 2},...$  ].
$a_{1,m}(t)$, $a_{2,m}(t)$ $b_m(t)$ and $c_r(t)$
are real coefficients and they contain the $t$ dependence
of the oscillatory modes.  $Q(t)$ is the $q+pt$ of (B.1) and
denotes the zero frequency mode.

Upon substituting the expansions (B.8) and (B.9) into the action (2.14),
we obtain,
as expected, the action $S$ and the Lagrangian $L$ for an infinite number
of harmonic oscillators:
$$  S=\int dt \;L\;,  $$
where
\begin{eqnarray}
{1\over{ \pi\nu}}\;
L&=& 2\kappa^2 {\dot Q}^2 - (N_1^2+N_2^2)  \nonumber \\
& &+\sum_{m=1}^\infty \biggl( (\kappa^2 {\dot a}_{1,m}^2 - m^2 a_{1,m}^2)
      + (\kappa^2 {\dot a}_{2,m}^2 - m^2 a_{2,m}^2)
+ (\kappa^2 {\dot b}_{m}^2 - m^2 b_{m}^2)  \biggr)\;   \nonumber  \\
& &+\sum_{r={1\over 2}}^\infty (\kappa^2{\dot c}_{r}^2 - r^2 c_{r}^2)
 \;.
\end{eqnarray}
The dot here denotes time differentiation.

In the Hamiltonian formalism, the momenta conjugate to $Q$, $c_r$,
$a_{a,m}$ and $b_m$ are $4\pi\nu\kappa^2 \dot Q$,
$2\pi\nu\kappa^2 {\dot c}_{r}$, $2\pi\nu\kappa^2 {\dot a}_{a,m}$ and
$2\pi\nu\kappa^2 {\dot b}_{m}$ respectively,
and the nonvanishing Poisson brackets are given by
$$
\{Q,\;\dot Q\}={1\over{4\pi\nu\kappa^2}}\;,\quad
\{a_{a,m},\;{\dot a}_{b,n}\}=
{1\over{2\pi\nu\kappa^2}}\delta_{a,b}\delta_{m,n} \;,\quad
$$
\be
\{b_m,\;{\dot b}_{n}\}={1\over{2\pi\nu\kappa^2}}\delta_{m,n}\;,\quad
\{c_r,\;{\dot c}_{r'}\}={1\over{2\pi\nu\kappa^2}}\delta_{r,r'} \;.
\ee

Next we replace the real coefficients
$a_{a,m}$, $b_m$ and $c_r$ along with their velocities by
the complex variables $\tilde a_{a,m}$, $\tilde b_{m}$ and
$\tilde c_{r}$ defined by
\be
\tilde a_{a,m}= \sqrt{\pi\nu\kappa}\;(
\kappa {\dot a}_{a,m} -im a_{a,m}  )\;,\quad
\tilde b_{m}= \sqrt{\pi\nu\kappa}\;(\kappa {\dot b}_{m}-im b_{m} )
\ee
\be
\quad {\rm and}\quad
\tilde c_{r}= \sqrt{\pi\nu\kappa}\;(\kappa {\dot c}_{r}-ir c_{r})\;.
\ee
Their nonzero Poisson brackets are all given by
\be
\{\tilde a_{a,m},\;\tilde a^*_{b,n}\}=-im \delta_{m,n}
\delta_{a,b}\;,\quad
\{\tilde b_{m},\;\tilde b^*_{n}\}=-im\delta_{m,n} \quad{\rm and} \quad
\{\tilde c_{r},\;\tilde c_{r'}^*\}=-ir\delta_{r,r'}\;.
\ee
Here, we allow the index $m$ in
$\tilde a_{a,m}$ and $\tilde b_{m}$ to be negative with
$\tilde a_{a,-m}=\tilde a_{a,m}^*$ and $\tilde b_{-m}=\tilde b^*_{m}$.
We also allow the index $r$ in $\tilde c_r $ to be negative with
$\tilde c_{-r}=\tilde c_r^*$.

In terms of these variables, the Hamiltonian for the system is
\be
{\cal H}= 2\pi\nu\kappa^2 {\dot Q}^2
+\pi\nu (N_1^2 + N_2^2)  +  {1\over{2 \kappa}} \sum_{m\ne 0}
\biggl(\tilde a_{1,-m}\tilde a_{1,m} + \tilde a_{2,-m}\tilde a_{2,m}
+\tilde b_{-m}\tilde b_{m} \biggr) + {1\over {2\kappa}}
\sum_{r=-\infty}^\infty \tilde c_{-r}\tilde c_{r}\;.
\ee

The currents $J^a_\pm(\sigma)$ can be expanded according to
\begin{eqnarray}
\sqrt{{\pi\kappa}\over\nu}
J^a_\pm(\sigma)&=& \sqrt{\pi\nu\kappa}\;(N_a\pm\kappa \dot Q )
\mp {1\over 2}\sum_{m=1}^\infty \biggl(
(i\sqrt{2}\tilde a_{a,\mp m} -\tilde b_{\mp m})\;e^{im\sigma} \;+ \;h.c.
\biggr)  \nonumber\\
& &\pm {{(-1)^a}\over 2} \sum_{r={1\over 2}}^\infty
(i\tilde c_{\mp r} e^{ ir\sigma}\;+ \;h.c.)  \;,
\end{eqnarray}
where $h.c.$ denotes hermitean conjugate and time dependence has been
suppressed.

If we take the sums and differences of $J^1_\pm$ and $J^2_\pm$, the
resulting functions ($J^{sum}_\pm$ and $J^{dif}_\pm$ of Section 4)
are $2\pi$ and $4\pi$ periodic respectively, that is
they satisfy eqs. (4.4) and (4.5).
The expansion given here for the currents
must therefore be equivalent to the one we wrote down
in Section 4.  Furthermore, the basis of test functions
given in that Section must
be valid for this system.  These expectations are readily verified.
The correspondence between the
coefficients $ \alpha^\pm_n,\; \beta_{n\over 2}$ defined in Section 4 and
the coefficients
$\dot Q,\; N_a,\;\tilde a_{a,m},\;\tilde b_m ,\; \tilde c_r$
defined here is given by
$$
\alpha_0^\pm=\mp\nu\;(N_1+N_2\pm 2\kappa \dot Q )\;,\qquad
\beta_0=-\nu\;(N_2-N_1 )\;,\qquad
$$
$$
\alpha_m^\pm= \mp\sqrt{{\nu} \over {\pi\kappa}}\;\biggl(
 {i\over \sqrt{2}}  {\rm sgn} (m)\;(\tilde a_{1,m}+\tilde a_{2,m})
\pm\tilde b_m\biggr)
\;,  \qquad
\beta_m= -i\; {\rm sgn} (m) \;\sqrt{\nu\over {2\pi\kappa}}\;
(\tilde a_{2,m}-\tilde a_{1,m})  \;,
$$
\be
\beta_r=i \; {\rm sgn}(r)\; \sqrt{\nu\over {\pi\kappa}}\; \tilde c_r\;,
\qquad  m=\pm 1,\pm 2,...  \;,
\qquad r=\pm{1\over 2},\pm{3\over 2},... \;,
\ee
where  sgn$(m)={m \over{|m|}}$.

\bigskip
\centerline{
{\bf a.} $\kappa_1\nu_1=\kappa_2\nu_2$ }

{}From the definitions of $\nu_a$ and $\kappa_a$,
$\kappa_1\nu_1=\kappa_2\nu_2$ corresponds to
the situation where the ratio of the tension to the velocity is
the same for both
loops of the figure eight.  Thus, for example, with the
velocities and tensions identical on the two loops,
we can still allow for loops of different length.  We will also
assume that ${\kappa_1}\over {\kappa_2}$ is irrational for simplicity,
for then the modes iii) and iv) are absent.
When the velocities on the two loops are equal,
this condition implies that ${\ell_ 1}\over {\ell_2}$ is irrational.

Unlike in case {\bf b }, there now exist solutions of type v).
This is because if $2 k_{a,n}$ are integers, then
${{2 k_{1,n}}\over {2 k_{2,n}}}={{ \kappa_{1}}\over {\kappa_{2}}}$
is rational, contrary to assumption.   Eq. (B.6) which governs
type v) solutions reduces now to
$$ \tan{\pi k_{1,n}} + \tan{\pi k_{2,n}}=0\;.$$
It leads to $k_{1,n}+k_{2,n}=n\;$ where $n$ is a nonzero integer.
Therefore for the eigenfrequencies $\omega_n$ in $k_{a,n}=
\kappa_a\omega_n$, we have,
\be
\omega_n ={n \over {\kappa_1 + \kappa_2}}\;.
\ee

The expansion of the fields $\chi_a$ can be written as
\begin{eqnarray}
 \chi_a(\sigma,t)&=&Q( t) + N_a\sigma  +\sqrt{2}\sum_{m=1}^\infty
 a_{a,m}(t)\;\sin{m\sigma}
\nonumber\\
& & + \sqrt{2}  \sum_{n=1}^\infty  d_{n}(t)
\;\biggr\{   \cos{k_{a,n}\sigma} +
\tan{\pi k_{a,n}} \sin{k_{a,n}\sigma} \biggl\}\;\;.
\end{eqnarray}
Upon substituting the expansions into the action (2.14), we obtain,
$$
S=\int dt \;L\;,
$$
\begin{eqnarray}
{1\over{ \pi}}L&
=&\biggl( \sum_{a=1,2}\nu_a\kappa_a^2 \biggr)\;{\dot Q}^2
- \sum_{a=1,2}\nu_a N_a^2
+ \sum_{m=1}^\infty  \sum_{a=1,2}
 \nu_a(\kappa_a^2 {\dot a}_{a,m}^2 - m^2 a_{a,m}^2)
\nonumber\\
& &+\biggl( \sum_{a=1,2}\nu_a\kappa_a^2 \biggr)\; \sum_{n=1}^\infty
{\sec}^2 \pi k_{1,n}\; ( {\dot d}_n^2-\omega_{n}^2 d_n^2) \; \;,
\end{eqnarray}
and the dot again denotes time differentiation.

In the Hamiltonian formalism, the momenta conjugate to $Q$,
$a_{a,m}$ and $d_n$ are $2\pi (\sum_{a=1,2}\nu_a\kappa_a^2 )\;{\dot Q}$,
$\;2\pi\nu_a\kappa_a^2 {\dot a}_{a,m}$ and
$2\pi(\sum_{a=1,2} \nu_a \kappa_a^2) \;
{\sec}^2 \pi k_{1,n} \;{\dot d}_n$ respectively,
and the nonvanishing Poisson brackets are all given by
$$
\{Q,\;\dot Q\}={1\over{2\pi (\sum_{a=1,2}\nu_a\kappa_a^2)}}\;,\quad  $$
\be
\{a_{a,m},\;{\dot a}_{b,n}\}={{\delta_{a,b}~\delta_{m,n} }
\over{2\pi\nu_a\kappa_a^2}}
\;,\quad
\{d_n,\;{\dot d}_{m}\}={ {\delta_{n,m}}
\over{2\pi\;(\sum_{a=1,2} \nu_a \kappa_a^2)\; {\sec}^2 \pi k_{1,n}}}
\;.
\ee

Next we define the complex variables
$\tilde a_{a,m}$ and $\tilde d_{n}$ according to
\be
\tilde a_{a,m}= \sqrt{\pi\nu_a \kappa_a}\;(
\kappa_a {\dot a}_{a,m}-im a_{a,m}) \;\quad   {\rm and}\quad
\tilde d_{n}={{C_n}\over {\sqrt{2}}}
 \; ({\dot d}_{n} -i\omega_n d_{n} )  \;,
\ee
where
\be
C^2_n=2\pi(\kappa_1+\kappa_2)
\biggl(\sum_{b=1,2}\nu_b\kappa_b^2 \biggr)\;{\sec}^2 \pi k_{1,n}\;.
\ee
Here we let $m,n$ to be negative with
$\tilde a_{a,-m}=\tilde a^{*}_{a,m}$
and $\tilde d_{-n}=\tilde d^{*}_{n}$.
The nonzero Poisson brackets involving (B.22) are
\be
\{\tilde a_{a,m},\;\tilde a^*_{b,n}\}=
-im\delta_{m,n} \delta_{a,b}\;,\quad
\{\tilde d_{n},\;\tilde d_{m}^*\}=-in \delta_{n,m}\;.
\ee

In terms of these variables, the Hamiltonian for the system is
\be
{\cal H}
=   \pi\; \biggl( \sum_{a=1,2}\nu_a\kappa_a^2 \biggr) {\dot Q}^2 +\pi\;
\biggl(\sum_{a=1,2}  \nu_a N_a^2\biggr)
 +\sum_{a=1,2}  {1 \over {2\kappa_a}}
\sum_{m\ne 0} \tilde a_{a,-m}\tilde a_{a,m}
+{1\over{ 2(\kappa_1+\kappa_2)}}
 \sum_{n\ne 0} \tilde d_{-n}\tilde d_{n}\;.
\ee

The currents $J^a_\pm(\sigma)$ are now given by
\be
{1\over {\nu_a}}J^a_\pm(\sigma)=N_a \pm \kappa_a {\dot Q}
- {i\over{\sqrt{2\pi\nu_a\kappa_a}}}\sum_{m\ne 0}
\tilde a_{a,\mp m} e^{ im\sigma} \pm
{\kappa_a}\sum_{n\ne 0}
\;{{{ \tilde d}_{\mp n}} \over {C_n}}
\sec {\pi k_{a,n}}\;  e^{ ik_{a,n}(\sigma-\pi)}\;.
\ee

Unlike in case ${\bf b }$, the
sums and differences of $J^1_\pm$ and $J^2_\pm$ are not
periodic, although they do satisfy the periodic boundary conditions (4.1)
and (4.2).   Thus the test function basis given in Section 4
can not be used here.  A basis for the test functions
$\Lambda=(\Lambda_+^1, \Lambda_+^2,\Lambda_-^1,\Lambda_-^2)$
is instead obtained directly from eq. (B.26).
The test functions associated with the
constant modes $N_1$, $N_2$ and $\dot Q$ are
\ba
\Lambda^{(N_1)}&=&  {1\over {4\pi\nu_1}} \;(1,\;0,\;1,\;0) ,
\nonumber\\
\Lambda^{(N_2)}&=& {1\over {4\pi\nu_2}}\;(0,\;1,\;0,\;1)
\nonumber\\
\Lambda^{(\dot Q)}&=& {1\over {8\pi\nu_1\kappa_1}}\;(1,\;1,\;-1,\;-1)
\ea
respectively.  They are normalized such that eq. (4.11) is satisfied.
The test functions associated with the oscillating modes
$\tilde a_{1,m}$, $\tilde a_{2,m}$ and $\tilde d_{n}$ are
\ba
\Lambda^{(\tilde a_{1,m})}&=& i \;
 \sqrt{\kappa_1\over {8\pi\nu_1}}
(e^{im\sigma},\;0,\; e^{-im\sigma},0)
\nonumber\\
\Lambda^{(\tilde a_{2,m})}&=& i\;
 \sqrt{\kappa_2\over {8\pi\nu_2}}
(0,\;e^{im\sigma},\;0,\;  e^{-im\sigma})   \;,
\\
\Lambda^{(\tilde d_{n})}&=& {{C_n}\over{8\pi}}
\biggl({{e^{ik_{1,n}(\sigma-\pi)}}\over
{\kappa_1 \nu_1 \sec{\pi k_ {1,n}}}}
,\;  {{e^{ik_{2,n}(\sigma-\pi)}}\over
{\kappa_2 \nu_2 \sec{\pi k_{2,n}}}}
,\; -{{e^{-ik_{1,n}(\sigma-\pi)}}\over
{\kappa_1 \nu_1 \sec{\pi k_{1,n}}}}
,\; -{{e^{-ik_{2,n}(\sigma-\pi)}}\over
{\kappa_2 \nu_2 \sec{\pi k_{2,n}}}}
\;\biggr)     \;.
\nonumber
\ea
They also satisfy eq. (4.11).

\end{document}